\documentclass[aps,pre,letterpaper,twocolumn,superscriptaddress,floatfix]{revtex4}
\usepackage{graphicx,psfrag,amsmath,amssymb,amsfonts,bbm,latexsym,color,dcolumn,bm}

\begin{document}

\title{Enhancement of fusion reactivities using non-Maxwellian energy distributions}

\author{Ben I. Squarer}

\affiliation{Department of Physics and Astronomy, Dartmouth College,  6127 Wilder Laboratory, Hanover 03755, USA}

\author{Carlo Presilla}

\affiliation{Dipartimento di Matematica, Sapienza Universit\`a di
Roma, Piazzale Aldo Moro 2, Roma 00185, Italy} 
\affiliation{Istituto Nazionale di Fisica Nucleare, Sezione di Roma 1, Roma 00185, Italy}

\author{Roberto Onofrio}

\affiliation{Department of Physics and Astronomy, Dartmouth College,  6127 Wilder Laboratory, Hanover 03755, USA}

\begin{abstract}
We discuss conditions for the enhancement of fusion reactivities arising from different choices of energy distribution functions for the reactants. 
The key element for potential gains in fusion reactivity is identified in the functional dependence of the tunnellng coefficient upon the energy, ensuring 
the existence of a finite range of temperatures for which reactivity of fusion processes is boosted with respect to the Maxwellian case. 
This is shown,  using a convenient parameterization of the tunneling coefficient dependence upon the energy, analytically in the simplified 
case of a bimodal Maxwell-Boltzmann distribution, and numerically for kappa-distributions,  We then consider tunneling potentials progressively 
better approximating fusion processes, and evaluate in each case the average reactivity in the case of kappa-distributions. 
\end{abstract}

\maketitle

\section{Introduction}

The relevance of controlled nuclear fusion in the current context, to contain global warming and to mitigate geopolitical conflicts, 
has been extensively debated. While the gap between experimental demonstrations and commercial use of nuclear fusion is being 
progressively narrowed with projects like ITER currently under construction, and DEMO, in the middle of this century, there have 
been parallel efforts to discuss the possibility to enhance fusion cross-sections by exploiting the basic physics of tunneling and the 
possible presence of screening of the Coulomb barrier. Considering the extreme sensitivity of quantum tunneling to the details of the 
process, significant gains may be expected. Examples of proposals range from discussion of correlated states \cite{Vysotskii}, 
interference from superposition of plane waves \cite{Ivlev}, use of generalized Gaussian wave packets \cite{Dodonov1,Dodonov2}, 
shielding of strong electromagnetic fields \cite{Lv}, the effect of the hypothetical presence of strong scalar fields \cite{Zhang}, 
among the many proposals. Another sequel of proposals has been focused on the intrinsic 
three-dimensional nature of the confined plasmas, with the goal to enhance the reactivity by producing Maxwell-Boltzmann (MB) 
distributions with different temperatures along different spatial directions \cite{Harvey1986,Nath2013,Kolmes2021,Li2022,Xie2023a,Xie2023b}.

In this paper we discuss the impact of various choices of macroscopic states for the reactants, {\it i.e.} their energy distribution, 
on the resulting average reactivity. Preliminary discussions of this aspect can be found in \cite{Majumdar2016} using Dagum 
distributions, and in \cite{Onofrio}, in which the potential gain in using energy distributions with hard high-energy tails of the
so-called kappa distributions ($\kappa$-distributions in the following) -- already broadly used in space plasma physics and 
astrophysics \cite{Livadiotis2011,Nicholls1,Nicholls2,Livadiotisa} -- has been discussed by evaluating reactivities with empirically determined 
fusion cross-sections \cite{Bosch}. 
We extend here these considerations to analytically evaluated {\it ab initio} cross-sections, showing general features 
and discussing conditions under which gains are expected with respect to Maxwell-Boltzmann (MB) energy distributions. 
A recent paper is also exploring, on top of trapping anisotropies, $\kappa$-distributions in magnetically confined plasmas \cite{Kong2024}.

The paper is organized as follows.
In section II we first recall general properties of two classes of non-MB distributions, bimodal MB and $\kappa$-distributions. 
We discuss the presence of population excesses at low and high energy, and population depletion at intermediate energies 
with respect to a Maxwell-Boltzmann distribution. We then report, in Section III, average reactivities gains in an idealized case of 
tunneling coefficient dependence upon the energy. In section IV we discuss tunneling in the case of two barriers which are amenable 
to a complete analytic treatment, yet capturing some features of the more complex nuclear fusion case, the double square well 
and a generalized form of the Wood-Saxon potential. In the same section we also provide explicit examples of configurations,  
within these two classes of potentials, for which it is advantageous to use $\kappa$-distributions. We briefly comment 
on the impact for fusion reactions involving deuterium-deuterium and deuterium-tritium mixtures, by using empirical cross-sections 
already available in the literature.

In the conclusions we qualitatively 
comment on the potential relevance of these results in the context of magnetic confined fusion reactors. 
Two appendices, one on explicit calculations for the tunneling coefficient of the square well case, and another on the 
discussion of the convexity of the tunneling coefficient evaluated with the WKB (Wentzel-Kramers-Brilllouin) approximation 
in the case of two relevant barriers, complete the paper.

\section{Generalized energy distributions}

In practical settings and especially in hot plasmas, the energy distributions of the reactants is determined by classical 
statistical mechanics. We will discuss here two energy distributions more general than the MB energy distribution, 
namely a superposition of two MB distributions at different temperatures, and the so-called $\kappa$-distribution. 

\begin{figure*}
\includegraphics[width=0.48\textwidth]{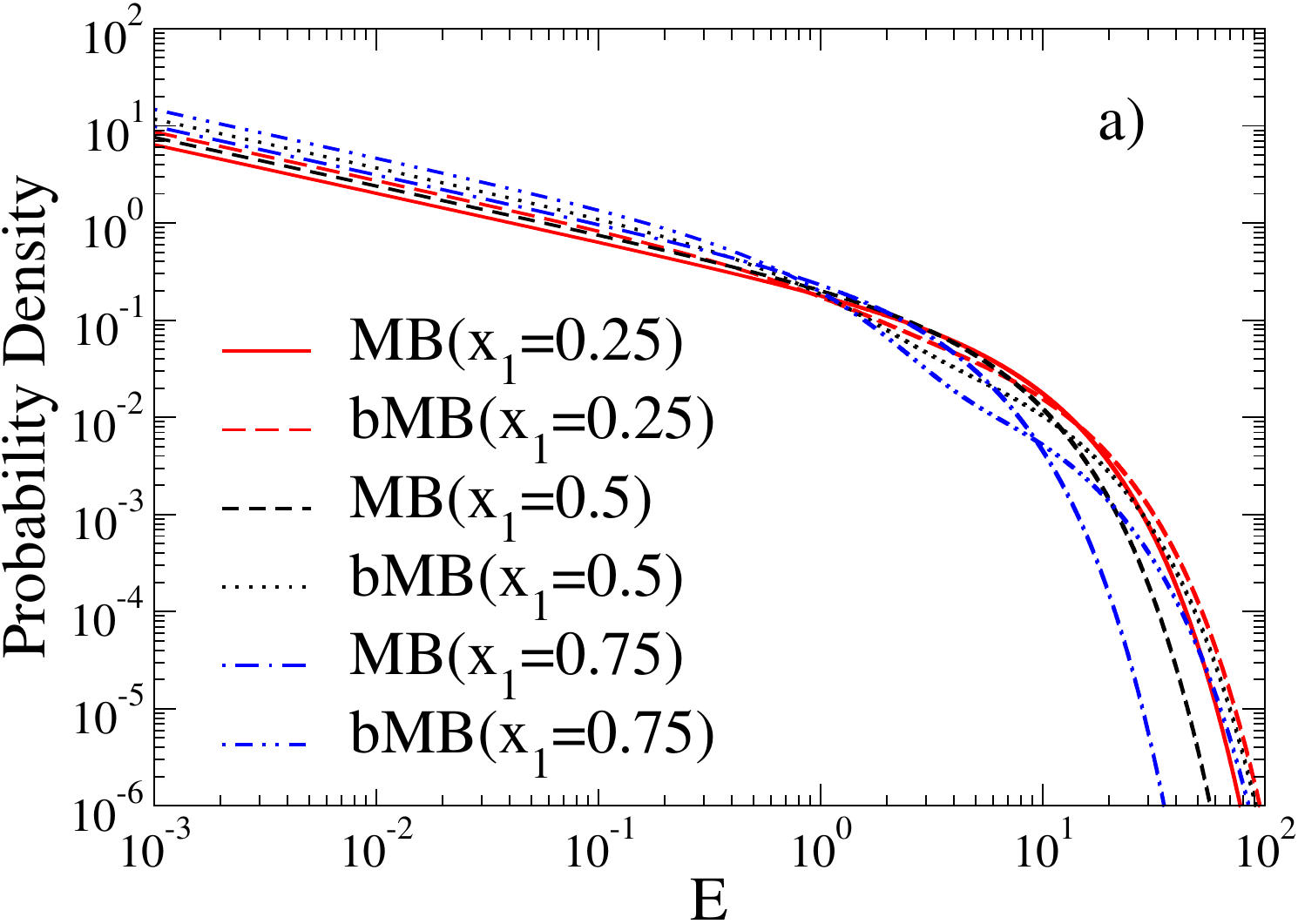}
\includegraphics[width=0.48\textwidth]{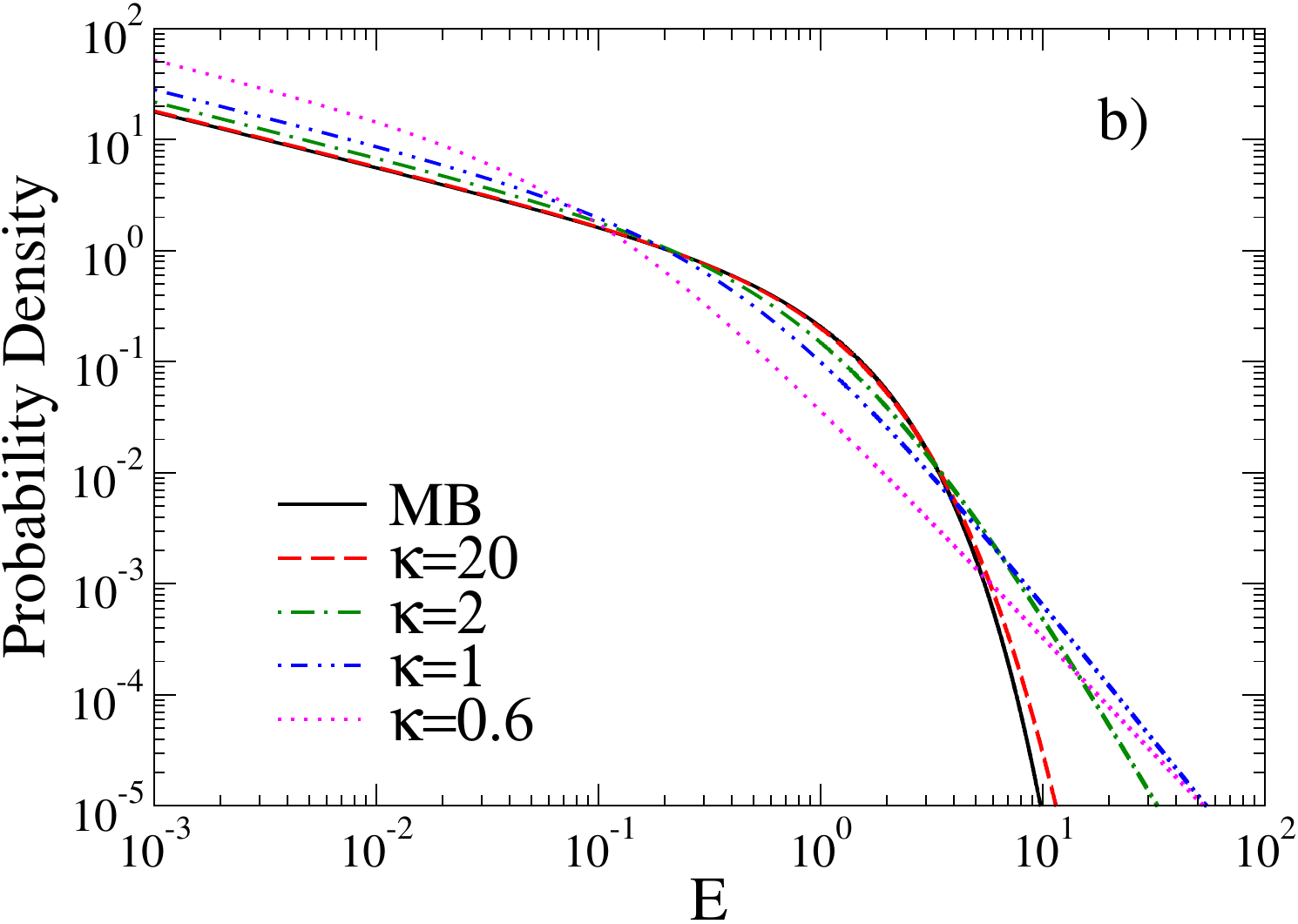}
\caption{Probability densities for non-Maxwell-Boltzmann energy distributions and for the corresponding MB energy distributions 
with the same average energy, versus $E$ (in arbitrary units).  
Three cases of bMB distributions are shown in (a), with inverse temperatures $\beta_1=1$ and $\beta_2=10^{-1}$ (in arbitrary units) for 
different weights of $x_1$=0.25,  with long dashed red (light grey) line, $x_1$=0.50 (dotted black line), $x_1$=0.75 (double-dotted-dashed blue line). 
The corresponding MB distributions with the same average energy and therefore inverse temperature $\beta_0$ (derived from 
Eq. (\ref{beta0})) are also shown  $x_1$=0.25 with continuous red (light grey) line, $x_1$=0.5 (short dashed black line), 
$x_1$=0.75 (blue dot-dashed line).
The case of $\kappa$-distributions is shown in (b) for different values of $\kappa=0.6, 1, 2, 20$, and constant $\eta=-1/2$.
For sufficiently large $\kappa$, the distribution approaches the MB distribution. A common feature of the two 
non-Maxwellian distributions with respect to the corresponding MB distribution is the presence of regions with 
higher probability densities at both low and high energy, with a region in between for which MB instead has higher probability densities.}
\label{fig1}
\end{figure*}

The MB in one dimension is defined as
\begin{equation}
p_{MB}(E;\beta)= \sqrt{\frac{\beta}{4\pi}} E^{-1/2} \exp(-\beta E),
\label{MB}
\end{equation}
with $\beta$, the unique parameter of this energy distribution, being the inverse temperature, such that the temperature $T$ 
is related to $\beta$ as $\beta=(k_B T)^{-1}$, with $k_B$ the Boltzmann constant, and $T$ expressed in Kelvin.
A bimodal Maxwell-Boltzmann distribution (bMB) is described by the weighted sum of two MB distributions
\begin{equation}
p_{bMB}(E; \beta_1, \beta_2)= x_1 p_{MB}(E; \beta_1)+ x_2 p_{MB}(E; \beta_2).
\label{bMB}
\end{equation}
Here the two weights $x_1$ and $x_2$ satisfy $x_1+x_2=1$, and the average energy of a bimodal system is 
$\langle E \rangle= x_1 \beta_1^{-1}+ x_2 \beta_2^{-1}$. 
We therefore can compare a bMB distribution to a single MB distribution with the same average 
energy, which means with a MB distribution with inverse temperature $\beta_0$ such that 
\begin{equation}
\beta_0^{-1}=x_1 \beta_1^{-1}+ x_2 \beta_2^{-1}.
\label{beta0}
\end{equation}
Examples of bMB energy distributions and comparison to the corresponding MB distributions with the inverse 
temperature determined by Eq. (\ref{beta0}) are shown in Fig.~\ref{fig1}a. Actual bMB distributions have been 
observed in laboratory plasmas \cite{Bhatnagar1993}.

The class of $\kappa-$distributions was introduced to fit magnetospheric electron data, and have been used to describe 
a plethora of astrophysical and space plasmas phenomena \cite{Livadiotisa}. 
It has been also discussed in the framework of nonextensive statistical 
mechanics in which the $q$-parameter, representing non-extensivity, is shown to be related to the $\kappa$-parameter \cite{Tsallis2009}.
Moreover, it has been shown that generalizations of bMB distributions allow to effectively capture the effect of a $\kappa$-distribution \cite{Hahn2015}.
In the one-dimensional case, $\kappa$-distributions can be expressed, as discussed in \cite{Vrinceanu}, in the following form 

\begin{eqnarray}
p_{\kappa}(E; \beta,\kappa,\eta)&=&
\sqrt{\frac{\beta}{4\pi (\kappa+\eta)}} \frac{\Gamma(\kappa+1)}{\Gamma(\kappa+1/2)} \times \nonumber \\
& & E^{-1/2} \left(1+\frac{\beta E}{\kappa+\eta}\right)^{-\kappa-1},
\label{Ekappa}
\end{eqnarray}
in which, in addition to the inverse temperature $\beta$, two further parameters appear with respect to a MB distribution, $\kappa$ and $\eta$. 
The parameter $\kappa$ expresses the `distance' of the distribution from the corresponding MB distribution, and 
determines more specifically the high-energy behaviour due to its presence in the generalized, power-law dependent, Lorentzian function.
Based on the usual definition of the exponential function as a limiting process $e^x= \lim_{n \to \infty} (1+x/n)^n$, it is evident that the 
MB distribution in Eq. (\ref{MB}) is recovered for $\kappa \rightarrow +\infty$.

For finite $\kappa$ instead the distribution has larger probability in the high energy tail with respect to the corresponding 
MB distribution, and these hard tails become more prominent as $\kappa$ becomes smaller. 
The average energy $E$ of a particle in a $\kappa$-distributed ensemble is written as 
\begin{eqnarray}
E&=&\frac{\kappa+\eta}{\beta} 
\frac{\int_0^{+\infty} x^{1/2}(1+x)^{-1-\kappa}}
{\int_0^{+\infty} x^{-1/2}(1+x)^{-1-\kappa}}dx= \nonumber \\
& & \frac{\kappa+\eta}{\beta} \frac{\Gamma(\kappa-1/2)}{2\Gamma(\kappa+1/2)}= \frac{\kappa+\eta}{2\beta(\kappa-1/2)},
\label{averageenergy}
\end{eqnarray}
where the dimensionless parameter $x$ is defined as $x=\beta E/(\kappa+\eta)$. Equation (\ref{averageenergy}) 
shows that the expectation of the equipartition principle valid for a MB distribution in one-dimension ($E=k_B T/2=1/(2\beta)$) is modified by a 
factor $(\kappa+\eta)/(\kappa-1/2)$, obviously tending to unity for $\kappa \rightarrow +\infty$, the MB limit. For the choice $\eta=-1/2$, the 
kinetic definition of temperature valid for a MB distribution is recovered regardless of the value of $\kappa$, allowing for a fair comparison 
between different $\kappa$-distributions therefore having the same total energy. 
This also coincides with the dependence of $\eta$ on the number of kinetic degrees of freedom $d_\kappa$ 
of the system (in our case $d_\kappa=1$), as $\eta=-d_\kappa/2$ \cite{Livadiotisb}. 
A comparison between $\kappa$ and MB distributions is also possible in general at the price however of introducing an effective 
temperature depending on $\kappa$.
For this reason, we will focus in the following considerations only on the simplest case of $\eta=-1/2$. 

Some remarks are also in order. First, the parameters $T$ and $\eta$ can be related to the velocity distribution's second moment
\begin{equation}
\langle v^2\rangle=\frac{\kappa+\eta}{\kappa-1/2} \frac{k_BT}{m},
\label{eq:variance}
\end{equation}
showing that both $T$ and $\eta$ are related to the distribution's variance (since the average velocity is zero), with the exceptional case 
of $\eta=-1/2$ commented above. 

\begin{figure*}
\includegraphics[width=0.48\textwidth, clip=true]{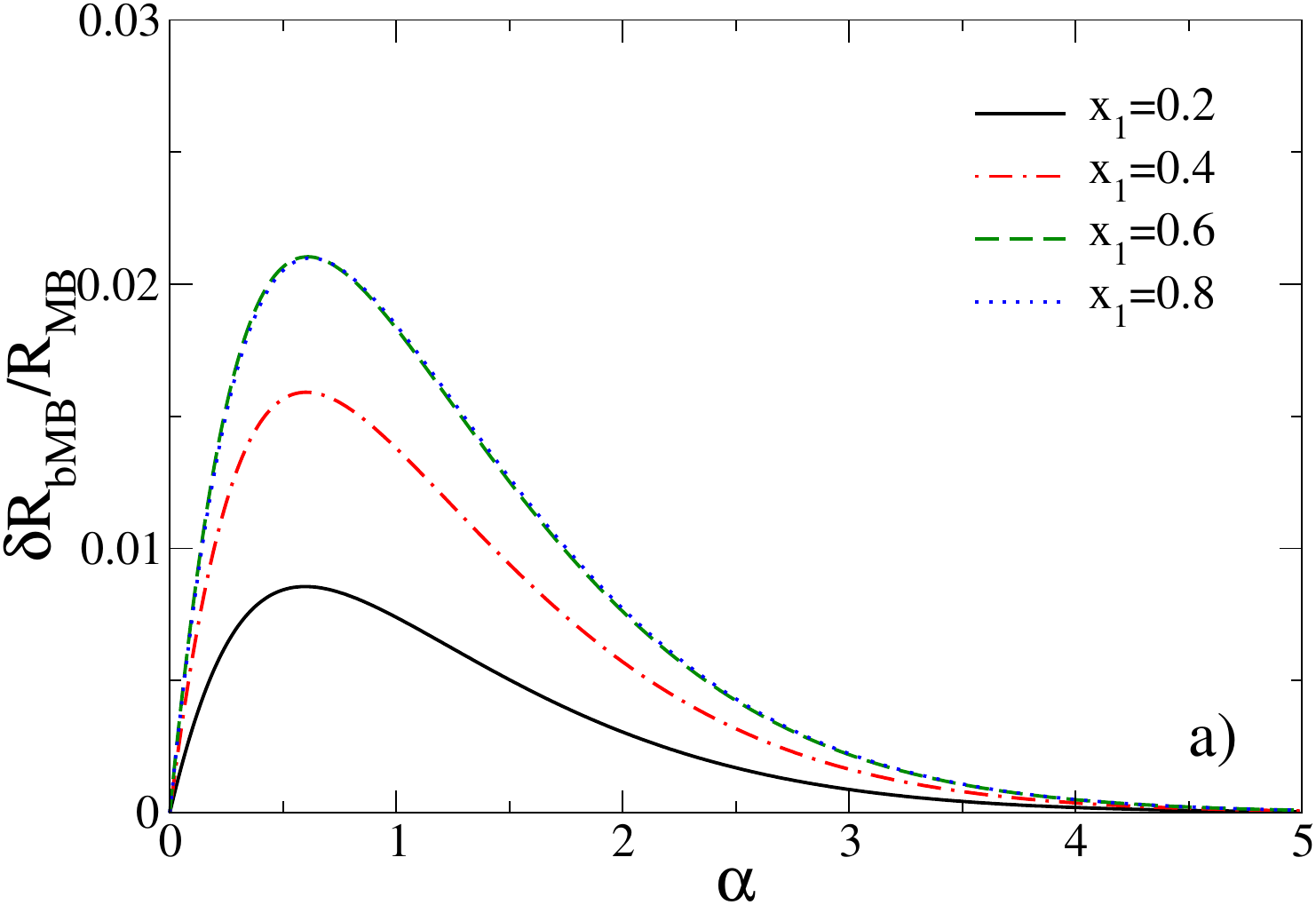}
\includegraphics[width=0.48\textwidth, clip=true]{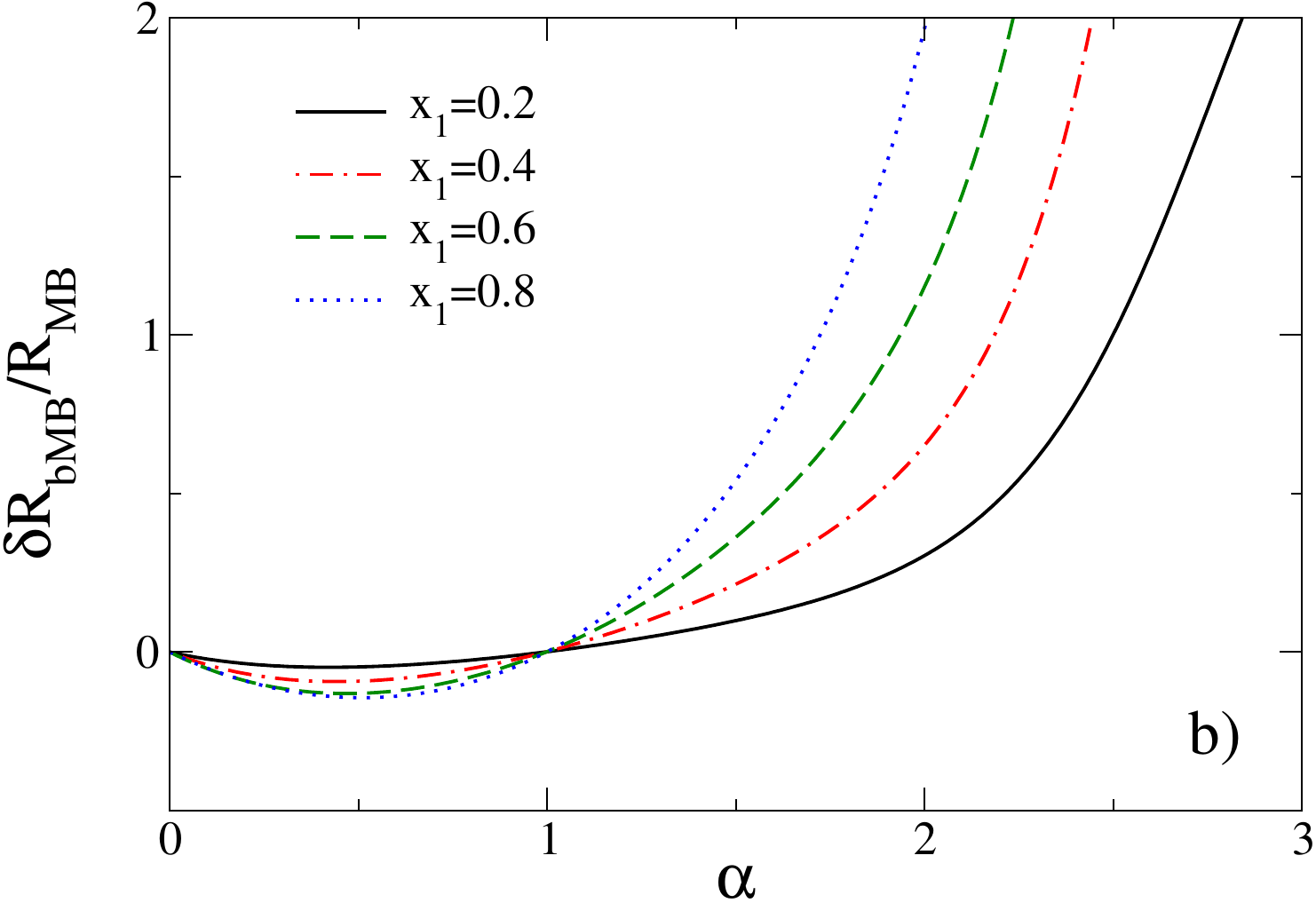}
\caption{Parameter $\delta R_{bMB}/R_{MB}$ versus the $\alpha$ exponent of the tunneling coefficient assumed in Eq. (\ref{Tunnelingalpha}).
Four cases of bMB with $x_1=0.2, 0.4, 0.6, 0.8$, all with $\beta_1=1, \beta_2=10^{-1}$, are depicted, and $\beta_0 E_0=10^{-2}$ (a), 
$\beta_0 E_0=10^2$ (b).}
\label{fig2}
\end{figure*}

Second, the velocity distribution $P_{\kappa}$ corresponding to the energy distribution $p_{\kappa}$ as in Eq. (\ref{Ekappa}), expressed as 
\begin{eqnarray}
P_{\kappa}(v; \beta,\kappa,\eta)&=&
\sqrt{\frac{m\beta}{2\pi (\kappa+\eta)}} \frac{\Gamma(\kappa+1)}{\Gamma(\kappa+1/2)} \nonumber \\
& & \times \left(1+\frac{\beta m v^2}{2(\kappa+\eta)}\right)^{-\kappa-1},
\end{eqnarray}
always reaches a maximum at $v=0$, and the ratio between the peaks of $\kappa$-distribution 
and the corresponding MB distribution for $v=0$ is
\begin{equation}
\frac{P_{\kappa}(v=0; \beta, \kappa,\eta)}{P_{MB}(v=0; \beta)}=\frac{\Gamma(\kappa+1)}{\sqrt{\kappa+\eta}\  \Gamma(\kappa+1/2)}.
\label{ratiof}
\end{equation}
By using asymptotic expressions for the Gamma-function, for instance a Stirling-like formula (see more in general \cite{Xu})

\begin{equation}
\Gamma(x+1) \sim \sqrt{2 \pi x} \left(\frac{x}{e}\right)^x,
\end{equation}
we find that in the case $\eta=-1/2$ we are considering, the ratio in Eq. (\ref{ratiof}) is always larger than unity for a finite $\kappa$, 
obviousy tending to unity in the $\kappa \rightarrow \infty$ limit. This implies that $\kappa$-distributions have both hard high energy tails 
and more populated peaks at zero velocity with respect to the corresponding MB distribution. Due to the normalization of 
probability distributions, this implies that there will be an intermediate regime of velocities in which the MB distribution 
prevails over the $\kappa$-distribution. This effect is shown in Fig.~\ref{fig1}b, in which various $\kappa$-distributions 
are considered including the limiting case of a MB distribution, nearly indistinguishable from a $\kappa$-distribution 
for $\kappa=20$. 

Third, an intriguing situation occurs in the limit of $\kappa \rightarrow -\eta$, as in this case, by introducing $\epsilon>0$ 
such that $\kappa=-\eta+\epsilon$, we have 

\begin{eqnarray}
P_{\kappa}(v; T, -\eta+\epsilon, \eta)&=& \sqrt{\frac{m}{2\pi k_BT}} \epsilon^{-1/2}
\;\frac{\Gamma(1-\eta)}{\Gamma(1/2-\eta)} \nonumber \\ 
& & \times \left(1+\frac{mv^2}{2 \epsilon k_BT}\right)^{\eta-1-\epsilon},
\end{eqnarray}
which, for $\epsilon \rightarrow 0$ and in the specific case of $\eta=-1/2$ we are considering, can be written as

\begin{equation}
P_{\kappa}(v; T, \eta, \epsilon) \simeq \frac{k_BT \epsilon}{mv^3},
\label{diverging}
\end{equation}
diverging in the limit of $v \rightarrow 0$ at finite $\epsilon$.
 
\section{Reactivity with non-MB distributions}

In this section we provide more quantitative arguments for understanding the effectiveness of the non-Boltzmann distributions with 
respect to a MB distribution (for previous related discussions see also \cite{Onofrio, Vrinceanu}) for processes involving tunneling 
phenomena, such as fusion. Before discussing the results, it is worth to comment on this specific, unusual situation, in 
which the velocity (and kinetic energy) distribution is dominated by classical physics, yet the reactants are evolving with fully quantum 
mechanical laws, either via Rutherford scattering (ineffective for nuclear fusion) or via fusion allowed by quantum tunneling. 
This implies that the variance of the single wave packet, so far considered as attributable to a Gaussian momentum distribution as 
customary for wave packets of quantum mechanical origin, is actually determined by the spreading of the velocities due to the 
classical distribution in the statistical ensemble, which in turn depends on temperature and density and the $\kappa$ and $\eta$ parameters
in the case of a $\kappa$-distribution (for a related discussion see \cite{Kadomtsev1997}). 
The cross-section corresponding to the tunneling process is

\begin{equation}
\sigma= \frac{\pi}{k^2} T(k)=\frac{\pi \hbar^2}{2 m E} T(E),
\end{equation}
where we have introduced the wave vector $k$ such that $\hbar k=\sqrt{2mE}$, and the velocity of the particle with wave vector $k$ is 
$v=\hbar k/m$. The average reactivity is then calculated as

\begin{eqnarray}
\langle \sigma v \rangle &=& \int_0^{+\infty} P(v) \sigma v dv =\nonumber \\ 
& & \frac{\pi \hbar^2}{\sqrt{2} m^{3/2}} \int_0^{+\infty} p(E) T(E) E^{-1/2} dE,
\label{Integrated}
\end{eqnarray}
with $p(E)$ one of the energy distributions defined in Eq. (\ref{MB}), Eq. (\ref{bMB}) or Eq. (\ref{Ekappa}). 

Basically the potential gain in using non-Boltzmann distributions stems from the fact that, with respect to a MB 
distribution, there is a more populated high energy tail. For the same average energy, this means that the non-Boltzmann distribution 
will also have a more populated low energy component, as visible in Fig.~\ref{fig1}. Then the advantage on using 
non-Boltzmann distributions relies on the functional dependence of the tunneling probability upon the energy, $T(E)$. 
If the latter is a convex function, the contribution to the integrated tunneling probability from the high-energy tail will overcompensate 
the lower contribution due to the increased component of the distribution at low energy. We also expect that at temperature large 
enough there will be marginal gain in using non-Boltzmann distributions. Indeed, for an arbitrary barrier we have $T \rightarrow 1$ 
for $E \rightarrow +\infty$, which implies that at high energy $T(E)$ curve will always be concave. We therefore mainly focus on 
the behavior at lower temperatures, which is also the most interesting region for fusion reactions of technological interest. 

\begin{figure*}
\includegraphics[width=0.48\textwidth, clip=true]{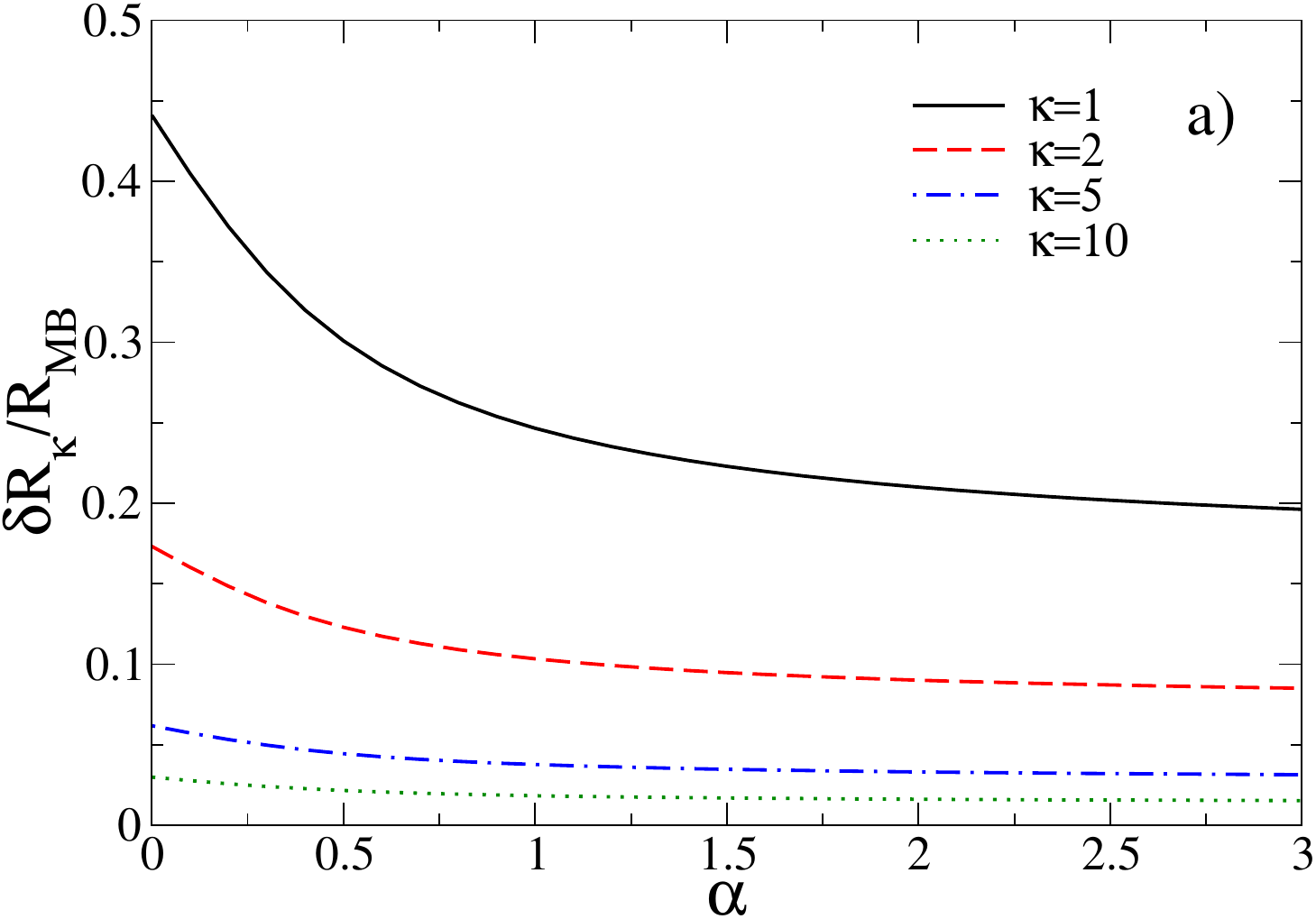}
\includegraphics[width=0.48\textwidth, clip=true]{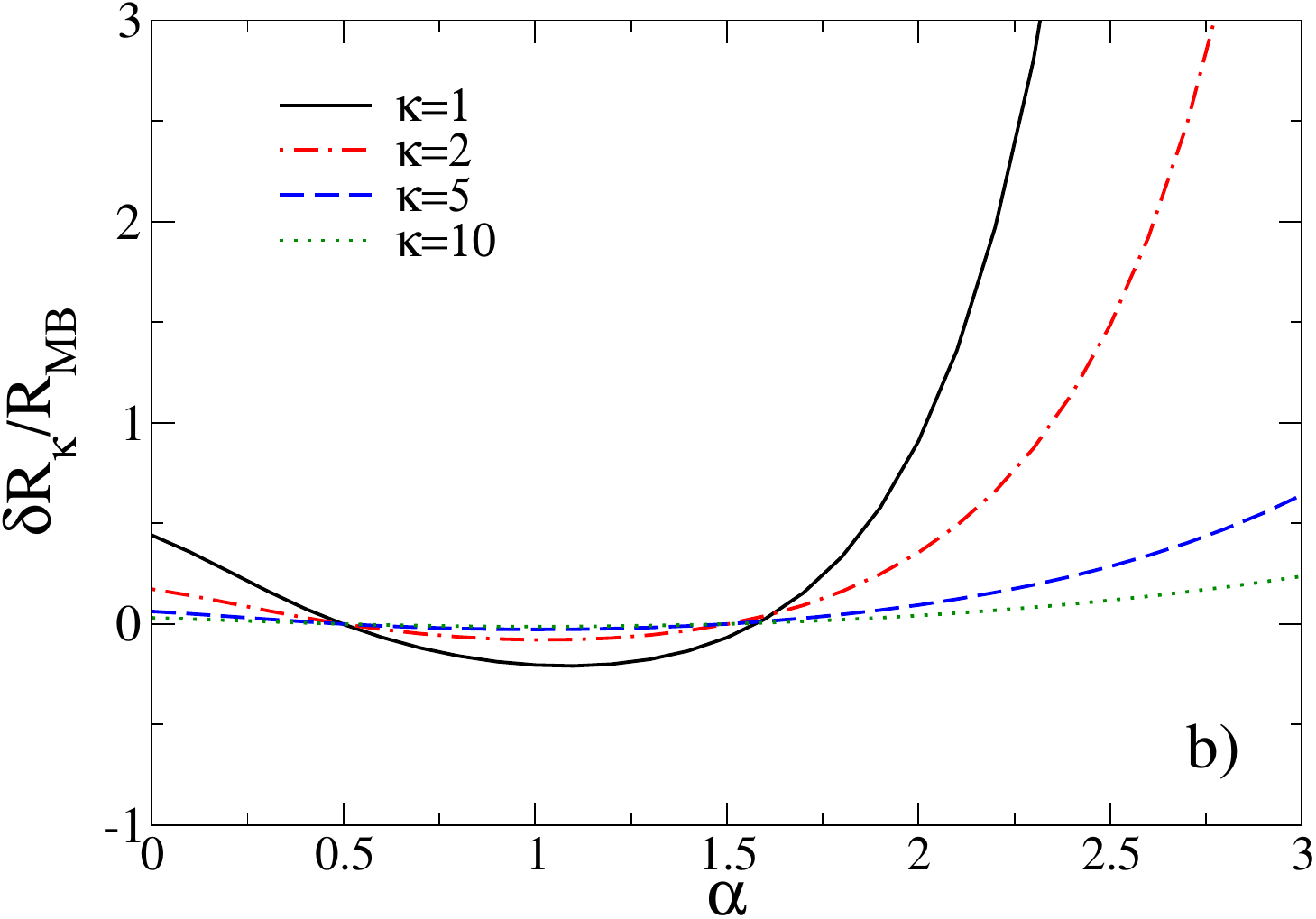}
\caption{Parameter $\delta R_{\kappa}/R_{MB}$ versus the $\alpha$ exponent of the tunneling coefficient.
Four cases are shown  with $\kappa=1, 2, 5, 10$ and $\beta E_0=10^{-2}$ (a), $\beta E_0=10^2$ (b). This plot differs 
from the one of bMB distributions as there is a gain in using $\kappa$ distributions also at very low $\alpha$, and 
because $\delta R_{\kappa}=0$ at two values of $\alpha$, at $\alpha=0.5$ independently of $\kappa$, and at a value of $\alpha$
progressively increasing with decreasing $\kappa$.}
\label{fig3}
\end{figure*}

For a generic non-MB distribution the gain with respect to the corresponding MB distribution may be quantified by considering 
the difference between the reactivities $\delta R$ defined as
\begin{eqnarray}
&& \delta R= \langle \sigma v\rangle - \langle \sigma v \rangle_{MB} = \nonumber \\ 
& & \frac{\pi \hbar^2}{\sqrt{2} m^{3/2}} \int_0^{+\infty}  [p(E)-p_{MB}(E)] T(E) E^{-1/2} dE.
\end{eqnarray}
A more practical and universal dimensionless parameter is obtained by considering $\delta R/R_{MB}$, the relative deviation from the 
MB reactivity. 

In a hypothetical case of $T(E)$ scaling exactly as $E^{1/2}$ in the entire energy range, this difference will be zero, as the two distributions 
are normalized to unity. However the tunneling coefficient cannot grow indefinitely, being limited to unity, so this scaling law does not allow for 
break-even in practice. The difference $p(E)-p_{MB}(E)$, for the two cases of non-MB distributions we are considering, 
is positive at low and high energy, being instead negative in a regime of intermediate energies, as already commented.
However, at low energy the tunneling coefficient is small, while at high energy the cross-section is small as evident 
by the explicit $E^{-1/2}$ factor, with the tunneling coefficient approximating unity. 

In order to put the discussion on a more quantitative ground, we introduce a fictitious tunneling coefficient depending on the energy as 

\begin{equation}
T(x)= \begin{cases} (E/E_0)^\alpha    & E < E_0 \\ \, \, 1  & E > E_0. \end{cases}
\label{Tunnelingalpha}
\end{equation} 
The parameter $\alpha$ ($0 \leq \alpha < +\infty$) plays the role of a `convexity' parameter such that for $E<E_0$ we have $T(E)$ 
convex if $\alpha>1$,  concave if $\alpha<1$. 

This allows to obtain a simple relationship for $\delta R/R$ in the case of bMB distributions. 
In this case the reactivity difference $\delta R$ is written as 
\begin{eqnarray}
\delta R_{bMB}&=& \langle \sigma v \rangle_{bMB} - \langle \sigma v \rangle_{MB} =\nonumber \\ 
& & \sqrt{\frac{\pi}{2}} \frac{\hbar^2}{m^{3/2}} (x_1 F_1+x_2 F_2-F_0),
\end{eqnarray}
where
\begin{eqnarray}
F_i(\beta_i; E_0, \alpha)&=&(\beta_i E_0)^{-\alpha} \Gamma(\alpha+1/2;0,\beta_i E_0)+ 
\nonumber \\ & & \Gamma(\alpha+1/2;\beta_i E_0,+\infty)
\label{Function}
\end{eqnarray}
with $i=0,1,2$, and
\begin{equation}
\Gamma(n;x,y)=\int_x^y e^{-t} t^{n-1} dt,
\end{equation}
is the incomplete Gamma function of order $n$. In Fig.~\ref{fig2} we show the plots of $\delta R_{bMB}/R_{MB}$ 
versus $\alpha$ for various values of $x_1$, for a bMB distribution. The case (a) is relative to a choice of 
$\beta E_0=10^{-2}$, case (b) deals with the opposite situation of $\beta E_0=10^2$. In the limit $\beta_i E_0 \rightarrow 0$, 
$F_i(\beta_i; E_0, \alpha) \rightarrow \Gamma(\alpha+1/2)$, the complete Gamma function, and then in the same limit 
there is complete cancellation among the three contributions, with the plot showing the small, residual term at finite 
$\beta_i E_0$. In the opposite limit,  $\beta_i E_0 \rightarrow +\infty$, the second term in the righthandside of 
(\ref{Function}) tends to zero, and then $F_i(\beta_i; E_0, \alpha) \rightarrow (\beta_i E_0)^{-\alpha} \Gamma(\alpha+1/2)$. 
This implies a limit form of the reactivity relative difference $\delta R_{bMB}/R_{MB}$

\begin{equation}
\frac{\delta R_{bMB}}{R_{MB}} \simeq \frac{x_1 \beta_1^{-\alpha}+x_2 \beta_2^{-\alpha}}{(x_1 \beta_1^{-1}+x_2 \beta_2^{-1})^\alpha}-1.
\end{equation}
It is easy to check that for $0 < \alpha <1$, $\delta R_{bMB}/R_{MB}<0$, and for $\alpha > 1$, $\delta R_{bMB}/R_{MB}>0$, with the 
bordeline case $\alpha=1$ yielding $\delta R_{bMB}/R_{MB}=0$. Therefore it is confirmed, at least for this power-law dependence of 
the tunneling coefficient upon the energy, that a convex function yields gain in using a bMB distribution with respect 
to a MB distribution with the same average energy. Considering the richness of possible situations, both in terms of 
$T(E)$ dependences, with the possibility of resonant tunneling for instance, and of possible non-Maxwellian distributions, this 
result has to be considered a qualitative guideline to appreciate the possibility of reactivity gains in a general context.

In the case of $\kappa-$distributions, we write an analogous relationship
\begin{eqnarray}
&& \delta R_{\kappa} = \langle \sigma v\rangle_\kappa - \langle \sigma v \rangle_{MB} = \nonumber \\ 
&&  \frac{\pi \hbar^2}{\sqrt{2} m^{3/2}}\int_0^{+\infty} [p_\kappa(E)-p_{MB}(E)]T(E)E^{-1/2} dE,
\end{eqnarray}
and the numerical integration provides the plots in Fig.~\ref{fig3} once again for choices of $\beta E_0=10^{-2}$ (a) and $\beta E_0=10^2$ (b).
Similarly to the bMB case, when $\beta E_0$ is small there is a small enhancement at all $\alpha$, due to the low-energy dominance 
of the $\kappa$ distribution with respect to the corresponding MB distribution. The case of large $\beta E_0$ is more interesting showing 
an interval of $\alpha$ values for which the reactivity of the MB distribution is significantly larger than the one of the $\kappa$ distribution, which 
instead prevails at small and large $\alpha$. The enhancement strongly depends on $\beta E_0$, and large values of this parameter, either a large 
$E_0$ or a low temperature $\beta^{-1}$, makes the reactivity quite sensitive to the high-energy tails, more prominent in the $\kappa$-distribution than in the MB case.
Also notice that, unlike the case of the bMB distribution, the threshold value of $\alpha$ for which the $\kappa$ distribution has higher reactivity occurs 
at $\alpha>1$, therefore requiring more curvature due to the presence of a strong low-energy component in the $\kappa$-distribution and a 
stronger depletion at intermediate energies, as noticeable by comparing the two panels of Fig.~\ref{fig1}.

\begin{figure*}
\includegraphics[width=0.48\textwidth, clip=true]{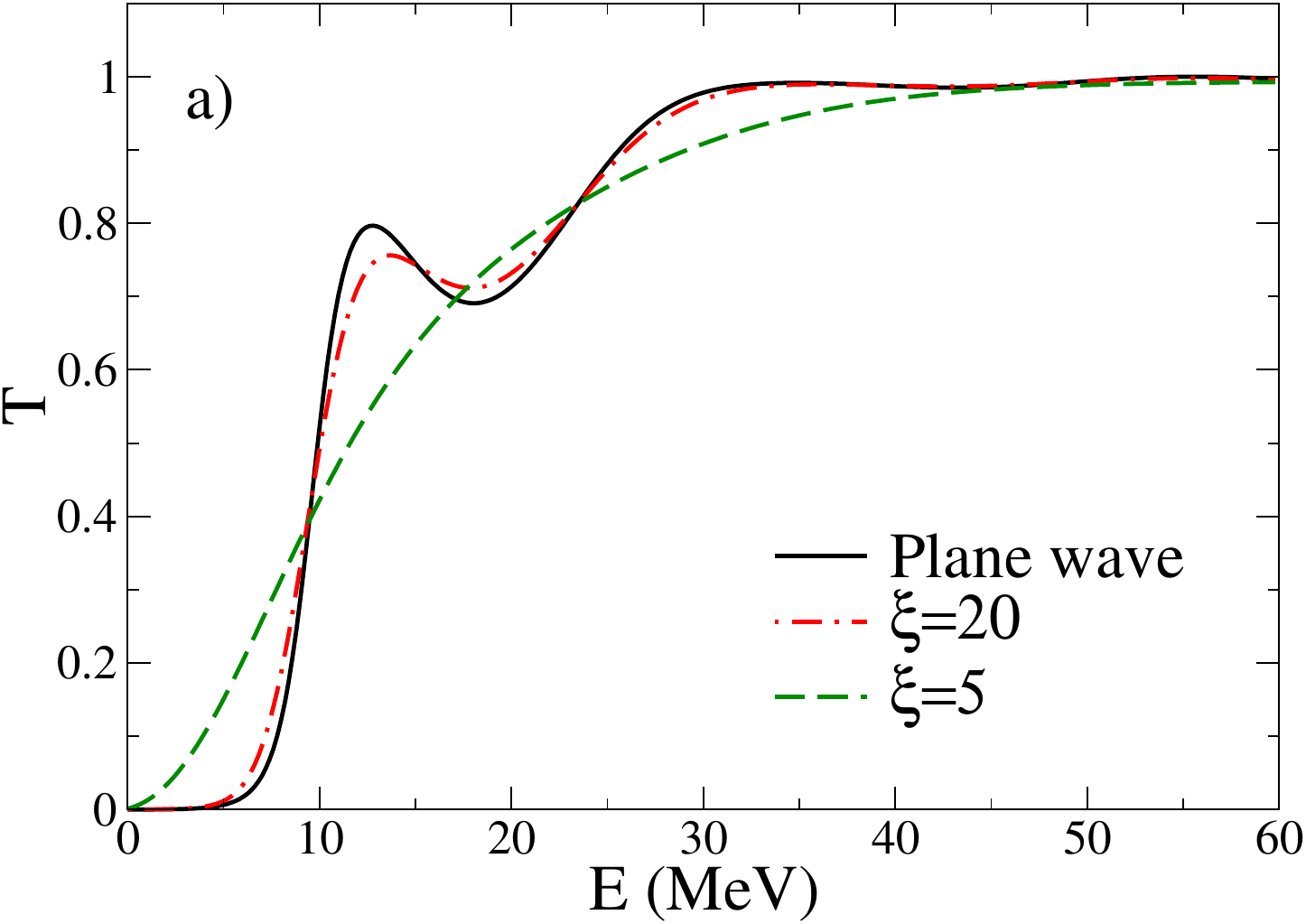}
\includegraphics[width=0.48\textwidth, clip=true]{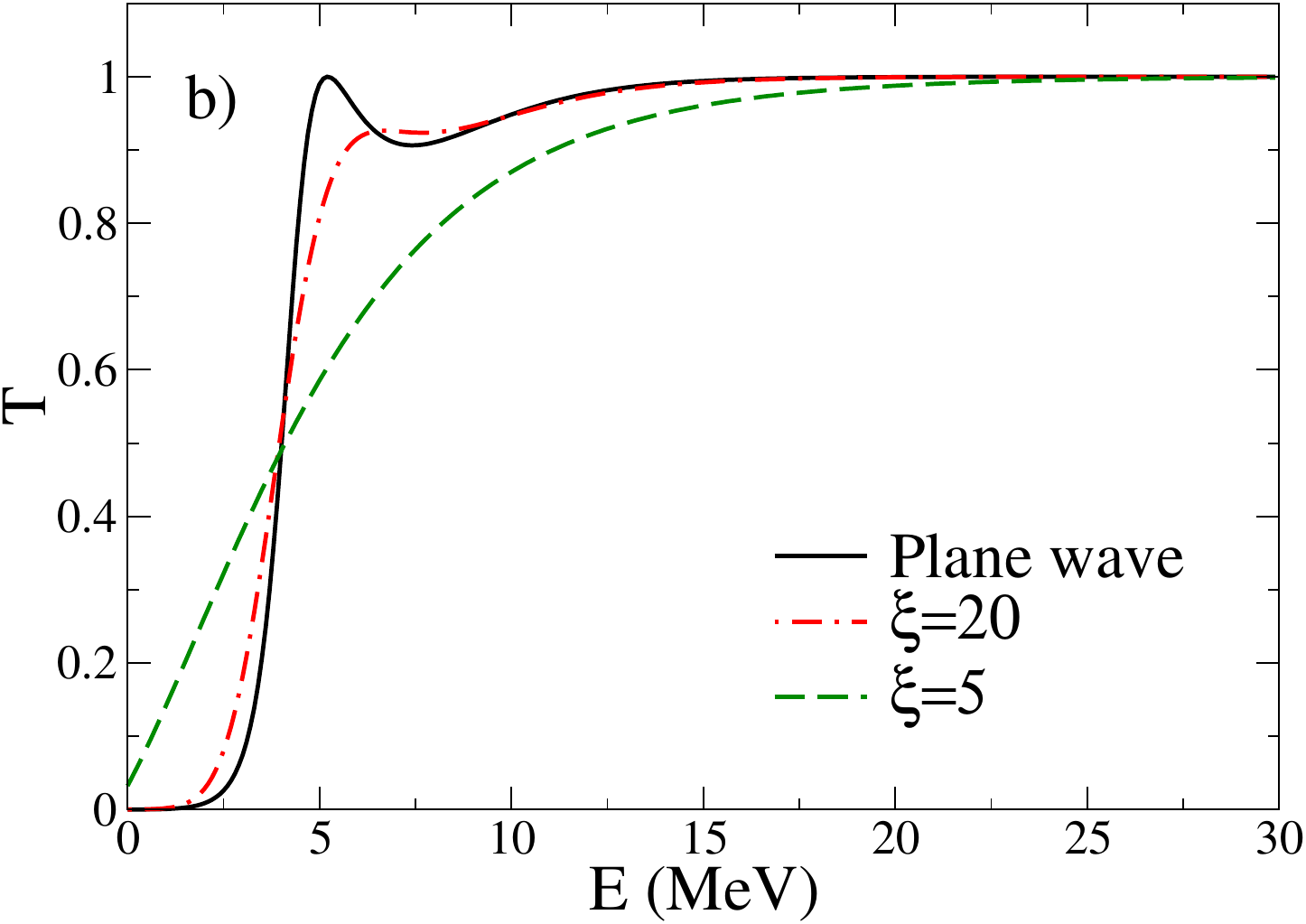}
\caption{(a) Tunneling coefficient versus the energy of the incident particle for a stepwise double barrier potential, with different cases of 
positional spreading of a Gaussian state, $\xi=5$ (green dashed line), $\xi=20$ (red dot-dashed line), and the case of a plane wave (black continuous line). 
The parameters of the potential are, with the notation used in Eq. (\ref{Potential}), $c=0.734, b=-c, d=c+5, a=-d, V_0=5.4, V_1=-40.33$.
(b) Same for the case of a GWS potential, with $\lambda=1.67, L=0.5, U_0=25, W_0=56$, all lengths being expressed in fm and all energies in MeV. 
The reduced mass is $m$=1 u.}
\label{fig4}
\end{figure*}

In the specific case of a $\kappa$-distribution with $\kappa \rightarrow -\eta$ discussed earlier, it is possible to support a similar 
convexity argument as follows. The reactivity in that case assumes the form
\begin{equation}
\langle \sigma v \rangle =\frac{\pi \hbar^2 k_B T\epsilon}{m} \int_0^{+\infty} \frac{T(v)}{v^4} dv,
\end{equation}
where $T(v)$ is the tunneling coefficient $T(E)$ expressed as a function of the particle velocity, $E=mv^2/2$, and we expect 
$T(v) \rightarrow 1$ for $v \rightarrow +\infty$. This means that the reactivity does not present divergences at large velocities. 
However there are possible divergences in the $v \rightarrow 0$ limit, depending on $T(v)$. Let us focus on the integrand 
at small $v$, including also the infinitesimal quantity $\epsilon$ for the analysis of the divergences. Let us suppose that 
$\epsilon \simeq v/v_0 \rightarrow 0$, {\it i.e.} goes to zero as the velocity, with $v_0$ a characteristic velocity, for instance the 
quadratic mean velocity. Then we will have an expression for the reactivity as
\begin{equation}
\langle \sigma v \rangle \simeq \frac{\pi \hbar^2 k_B T}{m v_0} \int_0^{+\infty} \frac{T(v)}{v^3} dv.
\end{equation}
Suppose that $T(E) \propto E^{\alpha}$ in the $E \rightarrow 0$ limit, then $T(v)  \propto v^{2\alpha}$, which means that the integral 
will be finite if $\alpha > 1$, and diverging at $-\infty$ otherwise. This confirms, within the limit of this example and related assumptions, that 
the $T(E)$ dependence should correspond to a convex function at least initially to avoid meaningless divergences of the reactivity. 
Notice that in this case the reactivity is directly proportional to the temperature.

 \section{Tunneling coefficients for potentials with analytic solutions of the Schroedinger equation}

In this section, we discuss tunneling probabilities for one-dimensional systems described by potentials progressively approximating the physical 
case of nuclear fusion but still admitting analytical solutions. We start by considering the case of a potential describing a stepwise double symmetric barrier,
then we discuss the most realistic case of a generalized Wood-Saxon potential. We investigate the tunneling phenomenon for different preparations 
of the wavefunction of the incident particle. As discussed in \cite{OnofrioPresilla}, there is a sensitive dependence of the tunneling coefficient upon 
the spatial spreading of the incident wavepacket. While we refer to this contribution for further details, we summarize here the results relevant for 
the current discussion. 

\begin{figure*}
\centering
\includegraphics[width=0.98\columnwidth]{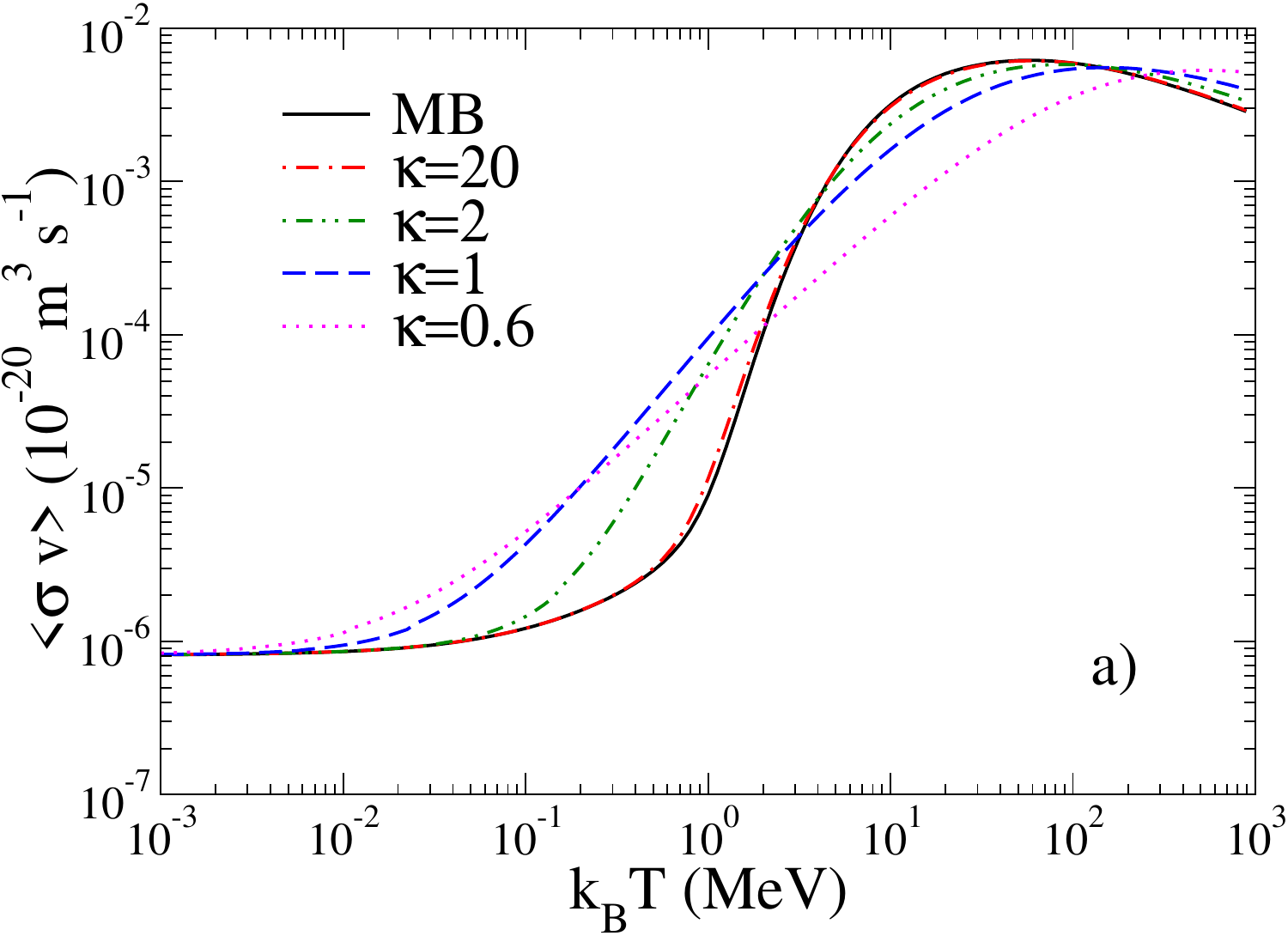}
\includegraphics[width=0.98\columnwidth]{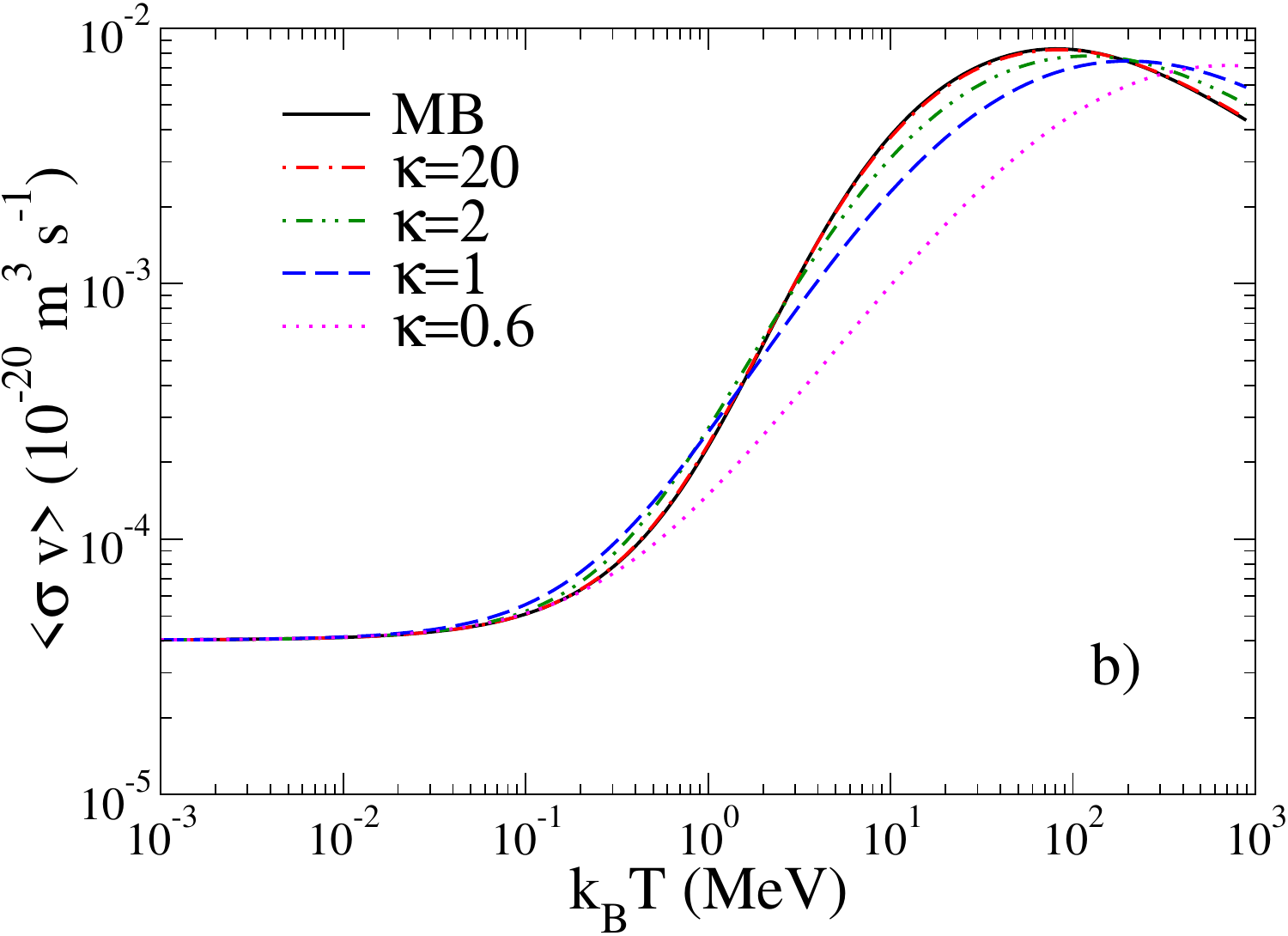}
\includegraphics[width=0.98\columnwidth]{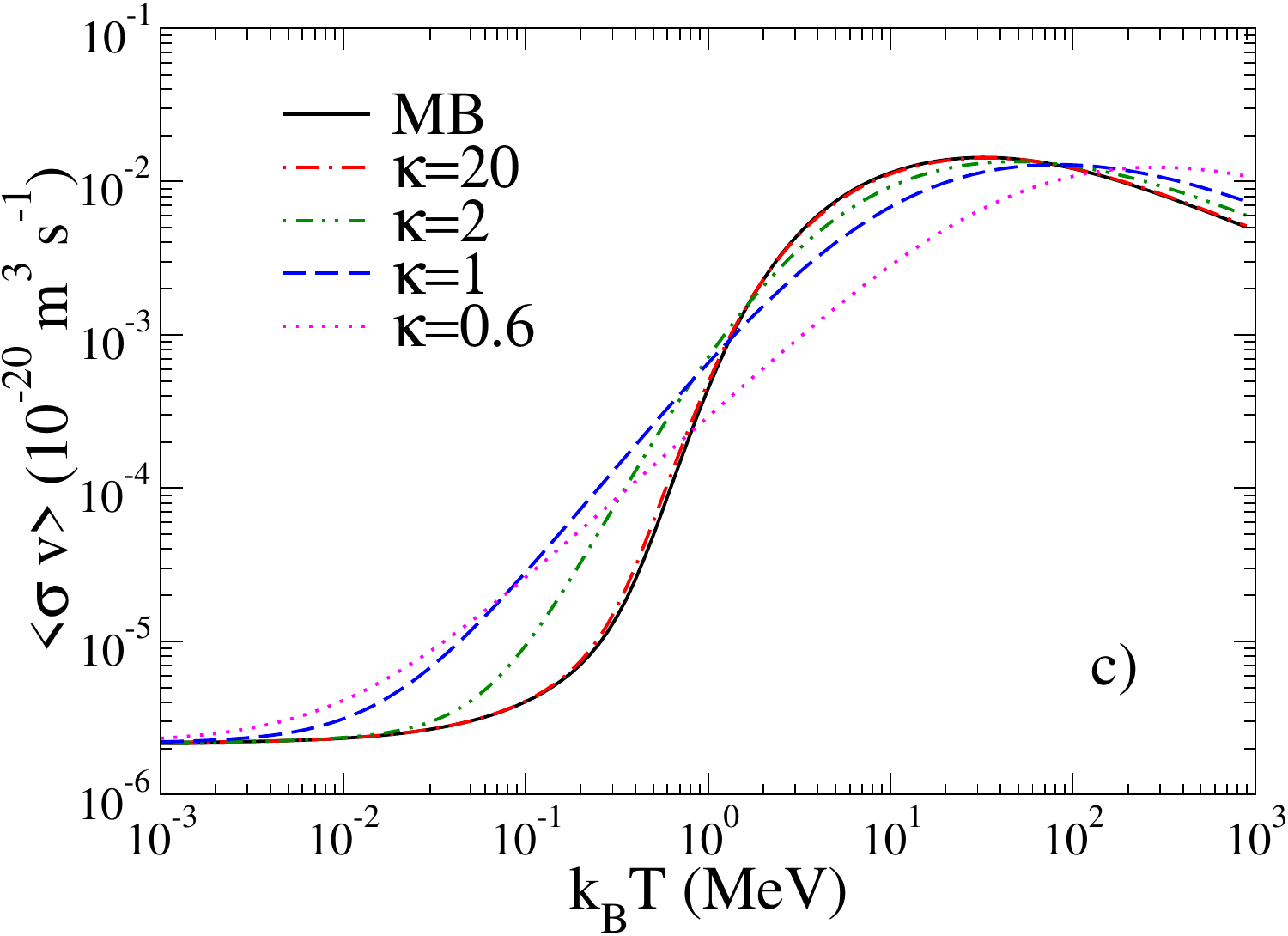}
\includegraphics[width=0.98\columnwidth]{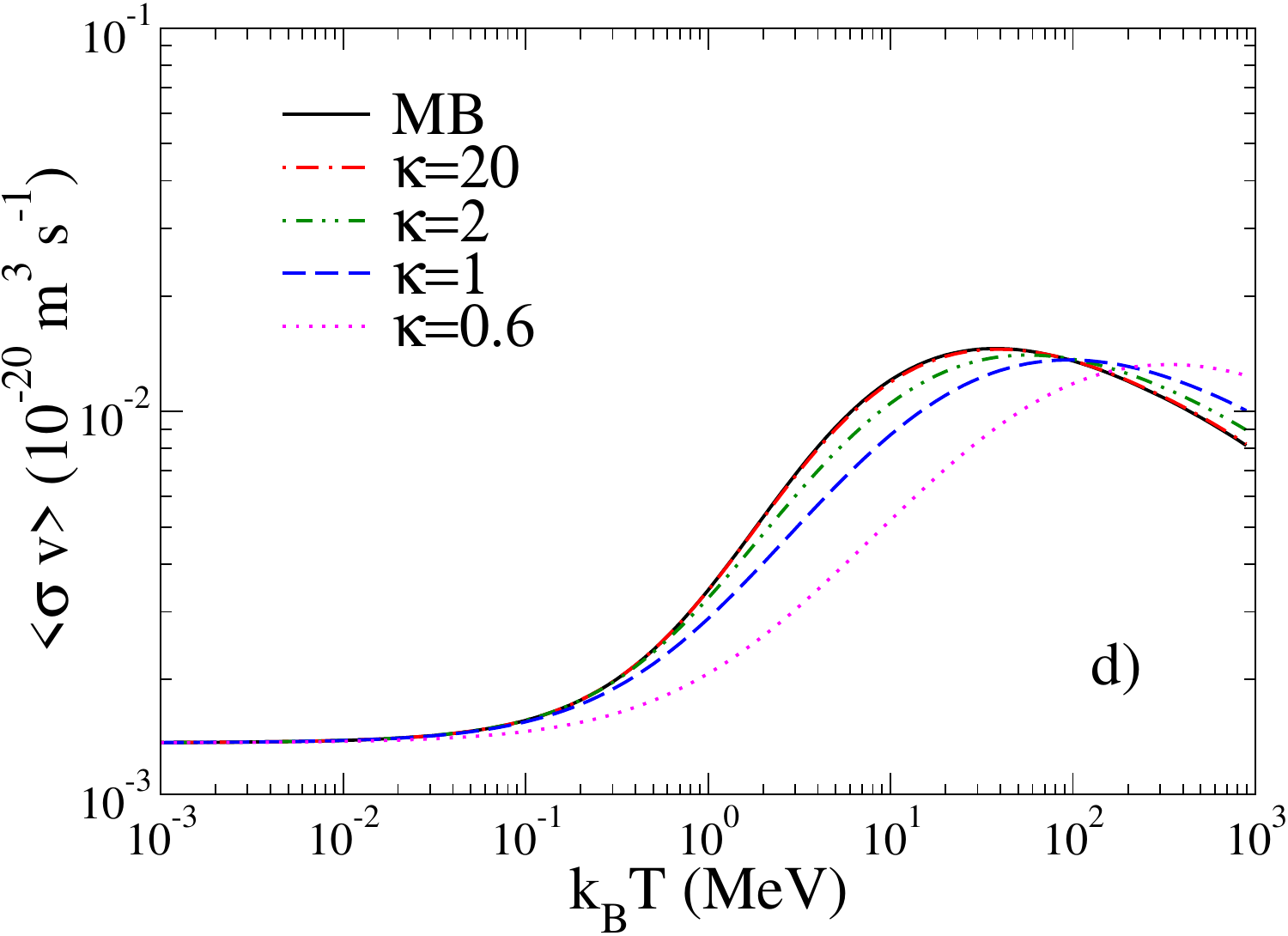}
\caption{Reactivity dependence upon temperature for different states and energy distributions. 
In the top panels we show the cases of a stepwise double barrier potential, with $\xi=20$~fm (a) and $\xi=5$~fm (b).  
In the bottom panels the corresponding cases of the GWS potential are shown for $\xi=20$~fm (c) and $\xi=5$~fm (d), respectively.  
The various cases in each plot are relative to the choice of $\kappa$-distributed energies, with $\kappa=0.6, 1, 2, 20$, and $\eta=-1/2$, and the 
corresponding MB distribution.}
\label{fig5}
\end{figure*}

The incident particle is schematized via a Gaussian wavepacket with positional spreading $\xi$ (such that the position variance is $\xi^2$), 
average wave vector $K$ and mean energy $\hbar^2 K^2/(2m)$:
\begin{equation}
  \psi(x,0)= \left(\frac{2}{\pi \xi^2}\right)^{1/4} e^{-(x-x_0)^2/\xi^2+i K x}.
\end{equation}
The corresponding wavefunction in wave vector space $k$ is
\begin{eqnarray}
  \varphi(k)
  &=&
      \frac{1}{\sqrt{2\pi}}
      \int_{-\infty}^{+\infty} \psi(x,0) e^{-i k x} dx
      \nonumber \\
  &=&
      \frac{1}{(2 \pi)^{1/4}} \sqrt{\xi}
      e^{-\xi^2 (k-K)^2/4} e^{i(K-k)x}.
\end{eqnarray}

Therefore the probability to measure a generic wave vector $k$ is a Gaussian
function of $k$ peaked around $K$
\begin{equation}
  P(k,K) = \vert \varphi(k)\vert^2
  = \frac{\xi}{\sqrt{2 \pi}} e^{-\xi^2 (k-K)^2/2}.
\end{equation}
The tunneling coefficient is therefore expressed by an integral over all wave vectors $k$ as
\begin{equation}
T(K)=\int_{-\infty}^{+\infty} dk~ T\left(\frac{\hbar^2 k^2}{2m}\right) P(k,K).
 \end{equation}
The reactivity is evaluated as

\begin{eqnarray}
\langle \sigma v \rangle_i&=&\frac{\pi \hbar^2}{\sqrt{2m^3}} \int_{-\infty}^{+\infty} dk
\int_{-\infty}^{+\infty} dK~ \sqrt{\frac{2m}{\hbar^2 k^2}} \times \nonumber \\ 
& & T\left(\frac{\hbar^2 k^2}{2m}\right)  P(k,K) P_i(K,\beta),
\end{eqnarray} 
where if the energy is $\kappa$-distributed ($i=\kappa$) we have

\begin{eqnarray}
P_{\kappa}(K,\beta)&=&\sqrt{\frac{\hbar^2\beta}{2\pi m(\kappa+\eta)}} \frac{\Gamma(\kappa+1)}{\Gamma(\kappa+1/2)} \times \nonumber \\
& & \left[1+\frac{\hbar^2 K^2}{2m} \frac{\beta}{\kappa+\eta}\right]^{-\kappa-1},
\end{eqnarray}
while in the MB case ($i$=MB) we have

\begin{equation}
P_{MB}(K,\beta)=\sqrt{\frac{\hbar^2\beta}{2 \pi m}} \exp{(-\beta \hbar^2 K^2/2m)}.
\end{equation}

\begin{figure*}
\centering
\includegraphics[width=0.48\textwidth]{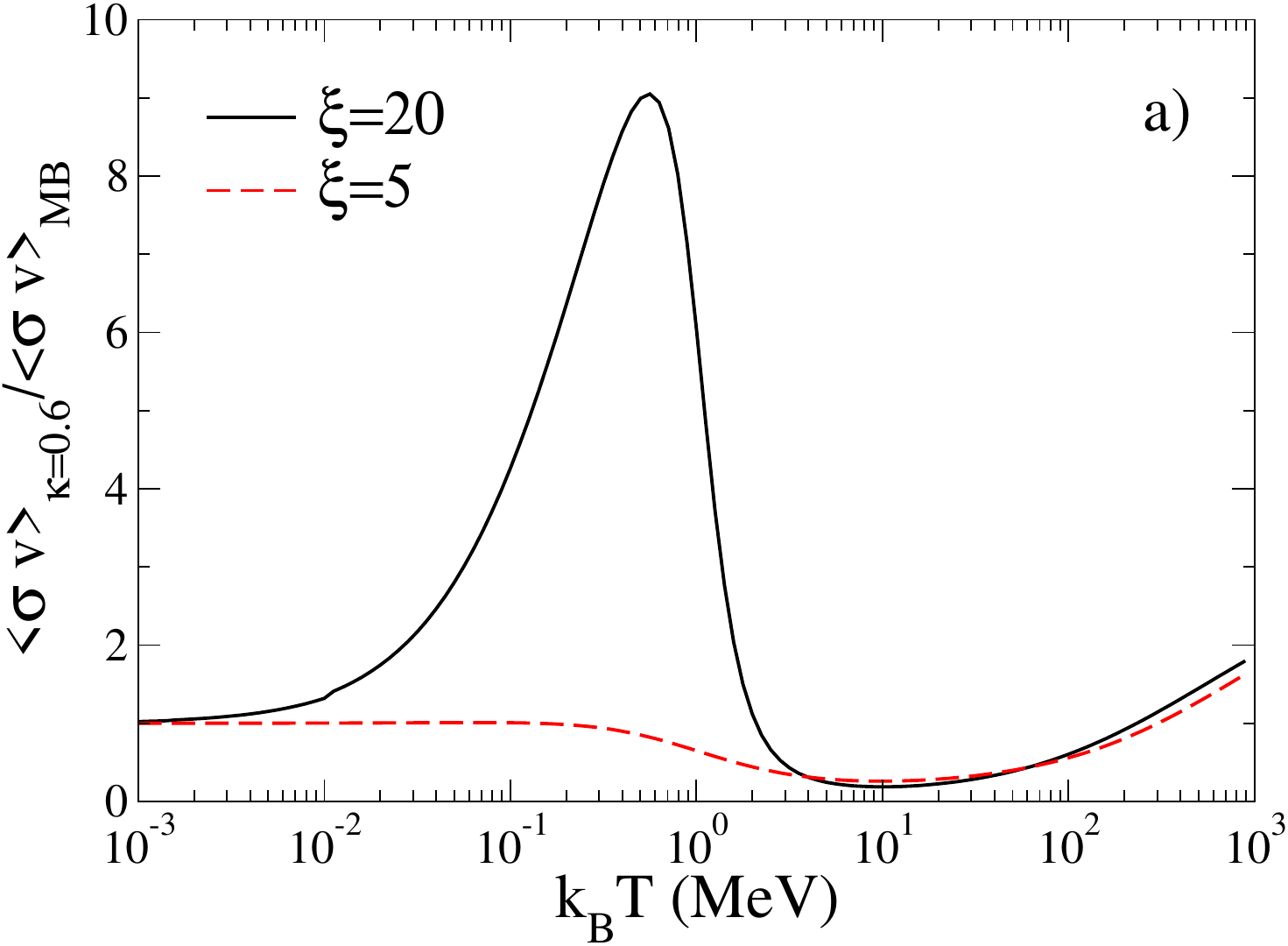}
\includegraphics[width=0.48\textwidth]{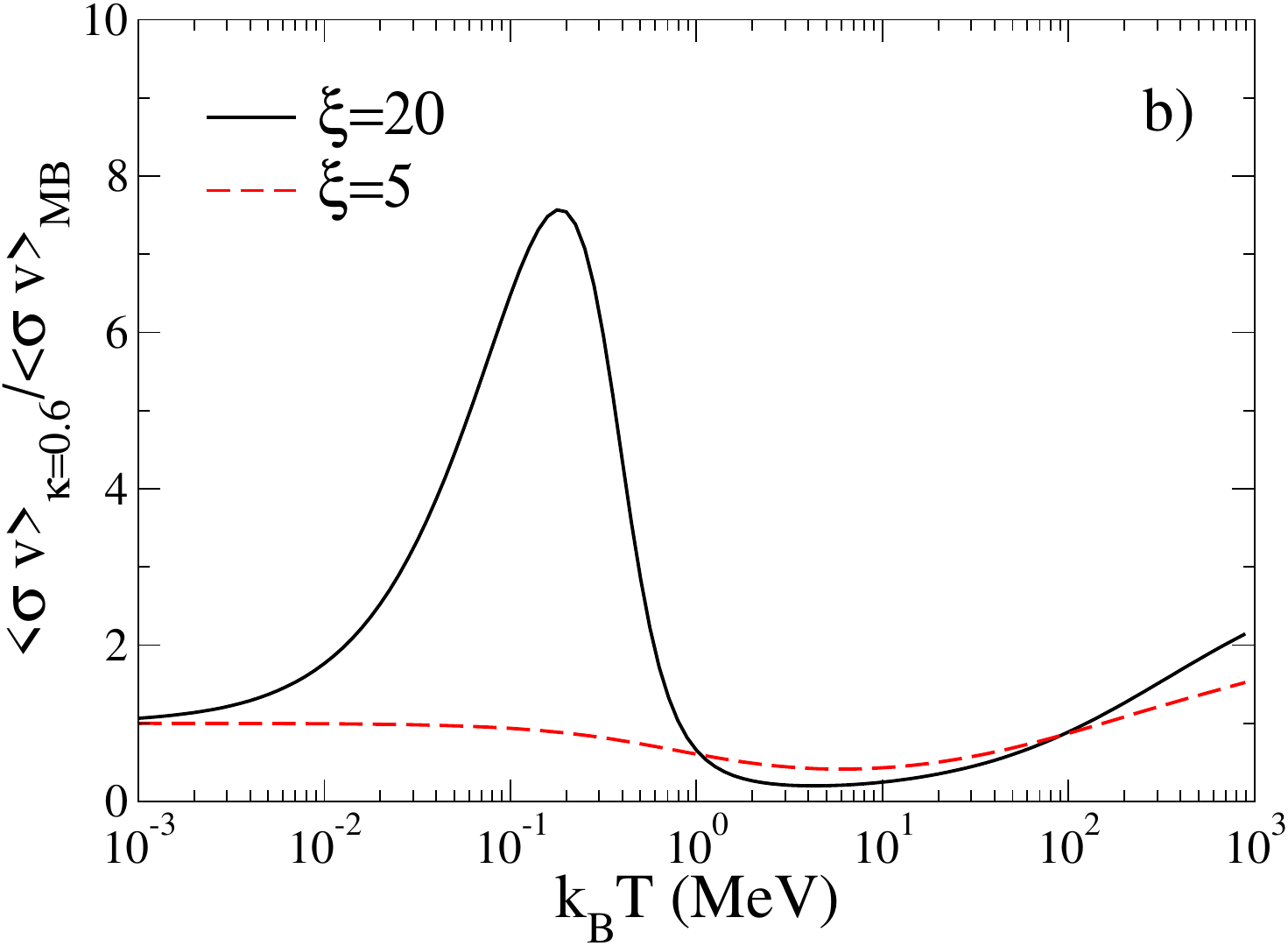}
\caption{Reactivity gain in using $\kappa$-distributed energies. The ratio of reactivities with the 
most favourable case of $\kappa=0.6$ and the case of a MB distribution is plotted for the two positional spreadings 
$\xi=5$~fm and $\xi=20$~fm, for the stepwise potential (a) and the GWS potential (b), same 
parameters as in Fig.~\ref{fig4}.}
\label{fig6}
\end{figure*}

As discussed in \cite{OnofrioPresilla}, the reactivity for fusion processes is extremely sensitive to the spreading of the Gaussian wavepacket, reaching a 
maximum for an intermediate value of $\xi$. We will discuss in the following both cases of highly localized Gaussian wavepackets as well as 
states resembling the limiting case of plane waves. Besides extending the analysis to localized states, we introduce now more realistic potentials leading to tunneling processes, in lieu of considering an artificial case as the one of Eq. (\ref{Tunnelingalpha}).  We first consider a stepwise double barrier potential defined as

\begin{equation}
V_{\mathrm{Step}}(x)= \begin{cases} 0     &  x < a \\
                 V_0              & a \leq x < b \\
                 V_1              & b \leq x < c \\
                 V_0              & c \leq x < d \\
                   0                & x  \geq  d  \end{cases}
\label{Potential}
\end{equation}              
where $a<b<c<d$ and $V_0$ e $V_1$ are both positive defined.
If $V_0 > V_1$ the potential $V(x)$ becomes a double barrier surrounding a well of depth $V_0 - V_1$.
This potential admits an analytic solution for the corresponding time-independent Schroedinger equation, and 
transmission and reflection coefficients are calculated as described in detail in Appendix \ref{AppA}. 
In plot (a) of Fig.~\ref{fig4} the dependence of the tunneling coefficient upon the energy of an incident particle is 
shown for two cases of positional spreading $\xi$, and of a plane wave. A distinctive feature of the various cases is that for 
large positional spreading the curvature of the tunneling coefficient is positive at low energy, {\it i.e.} $T=T(E)$ is initially a convex function. 
For small positional spreading the curve becomes concave in the whole range except in a tiny region near $E=0$. 

A more realistic case, at least because of the absence of discontinuities, is provided by the Generalized Woods-Saxon (GWS) 
potential energy for a one-dimensional system as first discussed in \cite{Luftuglu2016}
\begin{equation}
  V_{\mathrm{GWS}} (x) = -\frac{U_0}{1+e^{(|x|-L)/\lambda}} +\frac{W_0 e^{(|x|-L)/\lambda}}{(1+e^{(|x|-L)/\lambda})^2},
\end{equation}
parameterized by two characteristic lengths, $L$ and $\lambda$, and two energy scales, $U_0$ and $W_0$. 
The parameter $L$ determines the size of the effective well, and $\lambda$ is its spatial spread. 
The potential has a value in the origin equal to $-U_0/(1+\exp(-L/\lambda))+ W_0 \exp(-L/\lambda)/(1+\exp(-L/\lambda))^2$, while 
$-V_0/2+W_0/4$ at $|x|=L$.  At large distances $|x| \gg L$ the potential energy decreases as
$V(x) \simeq (W_0-V_0) \exp(-x/\lambda)$.  
This means that a semiqualitative difference from potential energies of interest in nuclear fusion is
that the barrier experienced by the nucleons, if schematized with this potential, does not have the long range 
as expected for Coulomb interactions, although in a realistic plasma the latter are screened on 
the Debye length. We choose the set of parameters as described in the caption of Fig.~\ref{fig4}, 
resulting in well depth, barrier height and width of the well comparable to the ones of light nuclei.

These potentials are reminiscent, in a one-dimensional setting, of the more general nucleus-nucleon interaction potential 
which also includes a Coulomb term inside the nucleus dictated by an assumed uniform electric charge density (here neglected) 
and a centrifugal term with the possibility for a scattering with non-zero impact parameter evidently absent in an one-dimensional analysis, 
see for instance \cite{Bekerman1988,Vanderbosch1992,Balantekin1998}.
The tunneling coefficient versus the energy of the incident particle is shown in plot (b) of Fig.~\ref{fig4} for different values of $\xi$. 
A similar phenomenon to the case of a stepwise double well is also visible, with the change of convexity depending on the values of $\xi$.
The presence of less defined boundaries with respect to the stepwise case makes resonant tunneling less remarkable especially 
in the $\xi=20$ case, with a barely visible peak around the energy of 6 MeV.

\begin{figure*}[t]
\centering
\includegraphics[width=0.48\textwidth]{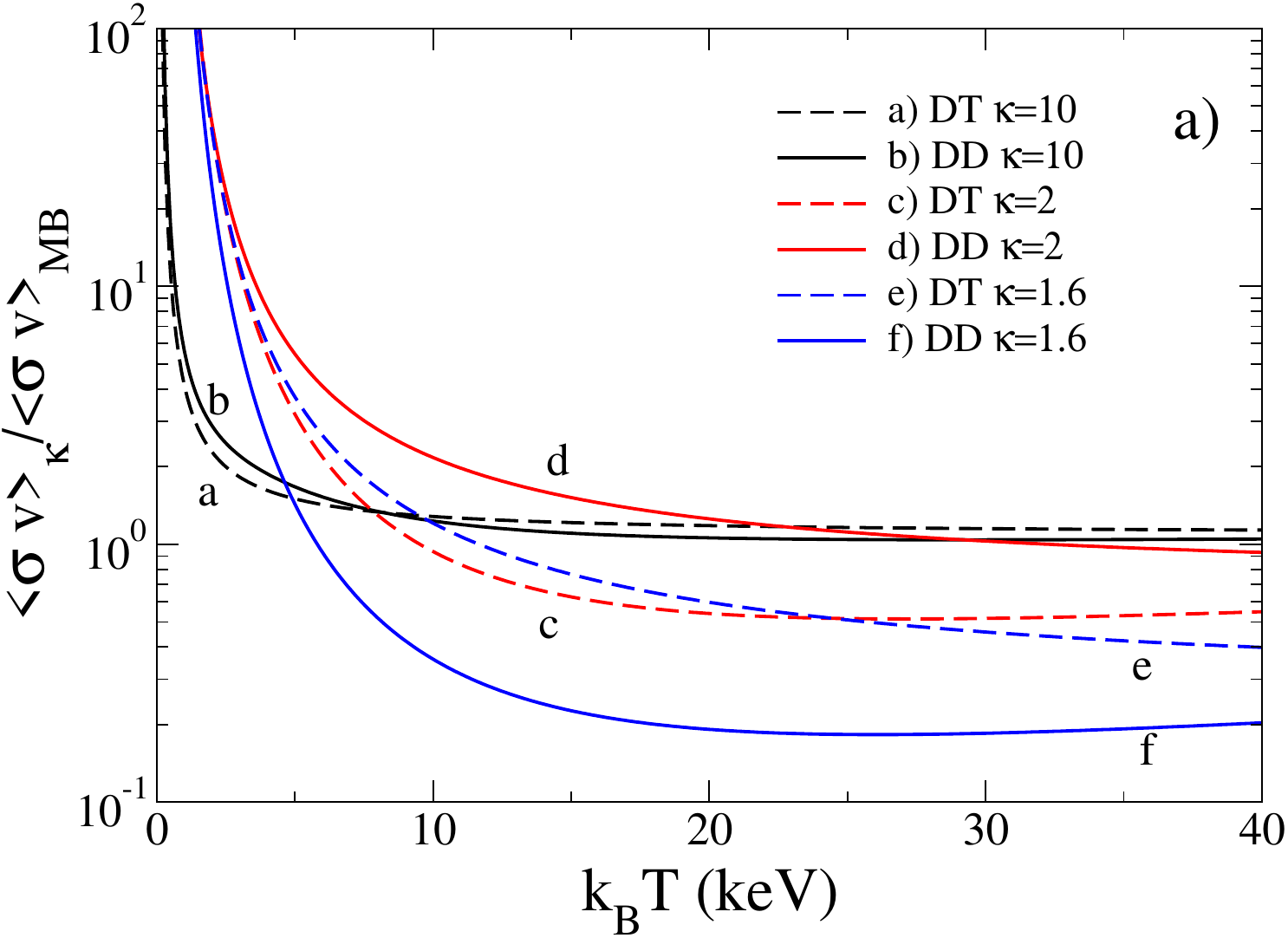}
\includegraphics[width=0.48\textwidth]{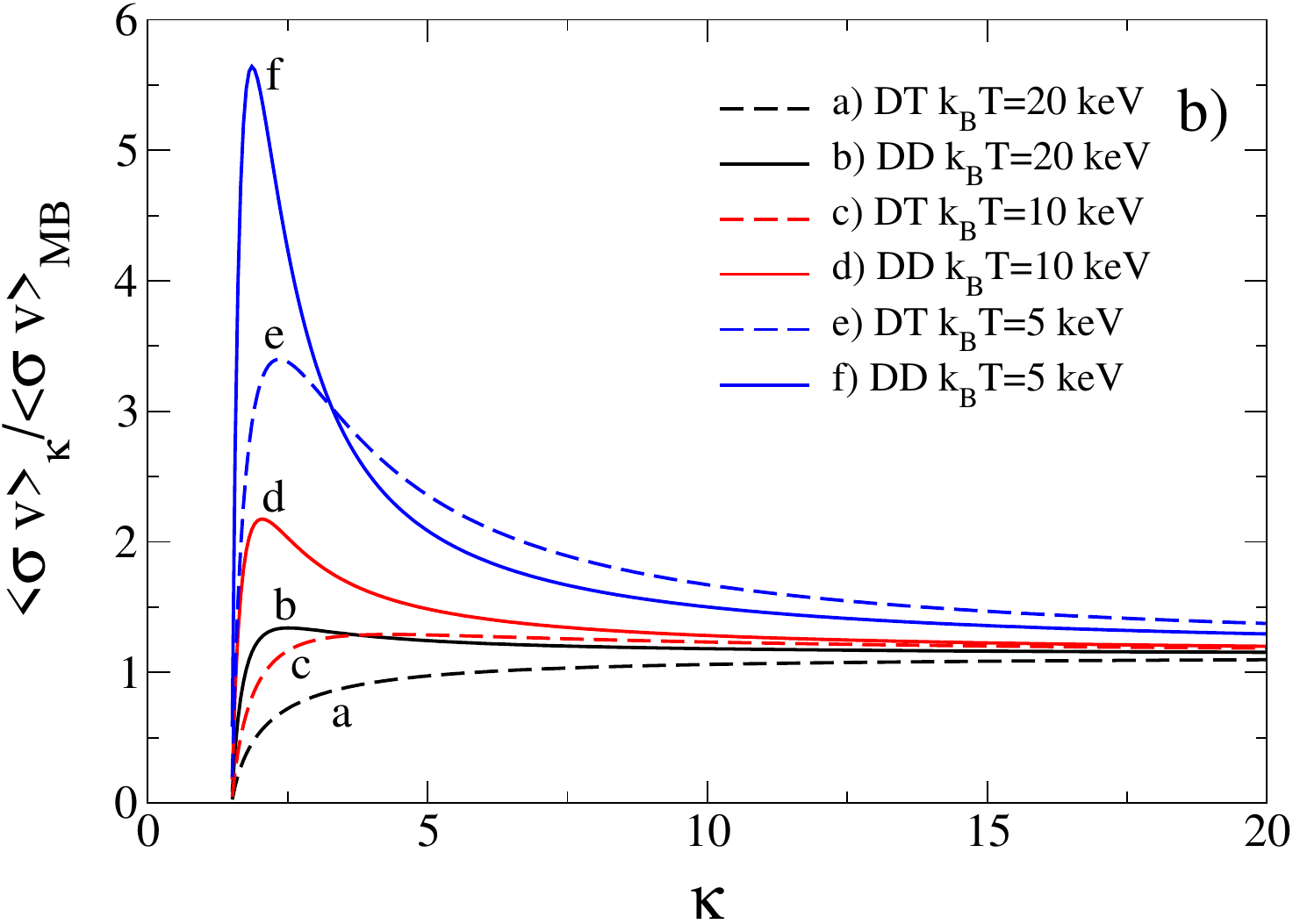}
\caption{Reactivity gain in using $\kappa$-distributed energies for deuterium-deuterium (continuous lines) and 
deuterium-tritium (dashed lines) plasmas. 
The ratio between reactivities with various $\kappa$-distributions ($\kappa=1.6, 2, 10$, with the $\kappa$=1.6 case very close to the 
threshold value having $\eta=-3/2$ in the three-dimensional case) and the case of a MB distribution 
is plotted versus temperature in a range of interest for the current experiments with Tokamak machines (a), and versus 
the $\kappa$-parameter for the three cases of temperatures of 5, 10, and 20 keV (b). 
The reactivities have been computed by using the experimental cross-sections parameterized as in 
\cite{Bosch}, by integrating the reactivity over energy intervals in which the interpolation is validated, 0.5-550 keV for the 
deuterium-tritium reaction, 0.5-5000 keV for D(d,p)T, 0.5-4900 keV for D(d,n)${}^3$He (see table IV in \cite{Bosch}).}
\label{fig7}
\end{figure*}

Based on these tunneling coefficients, we can evaluate the reactivities in some examples, for simplicity limiting the analysis to the case of $\kappa$-distributions.
Representative results for the reactivity dependence upon the temperature are shown in Fig.~\ref{fig5} for both stepwise and GWS potentials. 
As for the tunneling coefficients, cases of Gaussian states with narrow and broad positional spreading are considered. 
A broad Gaussian state, with the transmission coefficient having positive curvature at low energy as shown in Fig.~\ref{fig4}, at low 
temperature has a smaller reactivity for an MB distribution with respect to the corresponding $\kappa$-distribution. The opposite 
occurs in the case of a narrow Gaussian state in which the transmission coefficient has negative curvature in the entire range of energies. 
By increasing the temperature, the MB state dominates over the $\kappa$-distribution cases, until the harder energy tails of the latter determine once again 
a gain in reactivity at even higher temperatures. In order to better evidence the enhancement and suppression patterns, we show in Fig.~\ref{fig6} the ratio of 
reactivities between the case of $\kappa=0.6$, the most extreme $\kappa$-distribution we consider, and the case of the corresponding MB distribution. 
It is worth noticing that using broad Gaussian states with $\kappa$-distributions at small $\kappa$ allow for a gain, with respect to the MB distribution, 
of almost one order of magnitude, and most importantly in a range of temperatures below 1 MeV, of relevance for nuclear fusion. The analysis seems 
robust with respect to the choice of the potential energy, as shown by the similarity of the curves in the two cases considered. 

While we plan to discuss a comprehensive analysis of realistic cases of fusion in the future, it may be worth to briefly  
discuss the reactivity gain in using $\kappa$-distributions with respect to the MB case, evaluated from empirical cross-sections 
for fusion of deuterium-deuterium and deuterium-tritium mixtures \cite{Bosch}. In Fig. \ref{fig7} (a) this gain is plotted versus 
temperature in a range of interest for most of the experiments using Tokamak machines. 
The behavior is similar for the two mixtures. At large $\kappa$, as the $\kappa=10$ case, there is no gain at high temperature, 
while in the same temperature range it is not advantageous to use energy distributions with smaller values of $\kappa$. 
Significant gains are instead expected at lower temperatures. In this region of temperature there is an optimal, intermediate value of $\kappa$ maximizing 
the reactivity, since a too small $\kappa$ creates a nearly diverging population at very low energy -- see discussion around Eq. (\ref{diverging}) -- 
not favourable for fusion processes, see Fig. \ref{fig7} (b). Optimizing the value of $\kappa$ yields significant reactivity gains at low and already 
accessible temperatures.  These gains could be even substantial because in this analysis we do not consider the optimization with respect to the positional spreading. Moreover, the interval of energies for which the reactivity is evaluated is limited by the parameterization to a finite range, thereby resulting in 
conservative estimates, considering the importance of the high-energy tail for the more populated $\kappa$-distribution. This analysis, based on a concrete three-dimensional situation, also confirms the general trends reported earlier on more idealized and one-dimensional models.

\section{Conclusions}

By using generalizations of the Maxwell-Boltzmann statistics, we have discussed a key condition to enhance the 
tunneling probabilities and the reactivities of nuclear fusion processes in the framework of  one-dimensional potentials 
admitting exact solutions for the corresponding time-independent Schroedinger equation. 
A prominent convexity of the transmission coefficient ensures that the unavoidable overpopulation of low energy states 
does not offset the effect of the high-energy tail in a $\kappa$-distribution, resulting in enhanced integrated 
reactivity with respect to the Maxwell-Boltzmann energy distribution operating at the same effective temperature.
This has been explicitly shown for bMB and $\kappa$-distributions by using a convenient parameterization of 
the tunneling coefficient, and discussing how the latter can be approximated in more realistic, but still analytically 
tractable, potentials exhibiting tunneling. 

We have presented examples of potentials for which it is advantageous to use non-Boltzmann distributions to enhance fusion reactivity. 
The whole parameter space for which such an advantage exists can be explored via optimization of the reactivity ratio with respect to a 
few parameters, three in the case of the stepwise double barrier, and four in the case of the GWS barrier, 
for each temperature and positional spreading. The optimization is far more complex in terms of requested numerical resources 
by using a more realistic potential such as, for example, a combination of Wood-Saxon and Coulomb potentials.
It should be kept in mind that an overall optimization of the reactivity is what is beneficial to boost the fusion rate, therefore 
there is a competition between the choice of the positional spreading $\xi$ and the $\kappa$-parameter. As noticeable by comparing 
side by side the panels in Fig.~\ref{fig5}, even if the gain in using a small $\xi$ is smaller with respect to the choice of a large $\xi$ with 
the same $\kappa$ in the low temperature range, the absolute reactivities are larger in the former case.

Generalizations to more realistic cases requires handling tunneling coefficients and incorporating the possible 
presence of states with non-zero angular momentum in a full three-dimensional treatment, as well as the effect of 
the confining potential on the energy distribution. The extent to which $\kappa$-distributions can be realized in concrete 
setups is still open, however it is expected that long-range interactions in a generic statistical system will show 
deviations from a Maxwell-Boltzmann distribution rigorously valid only for short-range interactions. 
This point is extensively discussed in \cite{Tsallis2009} and corroborated by numerical simulations in 
the case of the Hamiltonian mean-field model \cite{Konishi,Antoni}.  Although we still lack evidence for 
$\kappa$-distributed energies in the case of genuine dynamical systems such as interacting classical gases, 
steps towards this direction are currently undergoing \cite{JauffredOnoSun2019,OnoSun2022}. On the experimental 
side, spectroscopy of the fusion reaction products is expected to provide precise assessments of the deviation from MB distributions, 
as recently discussed in \cite{Crilly2022}.

\appendix

\section{Tunneling coefficient in a stepwise double well potential}%
\label{AppA}

We report here the results for a potential made of a double well, starting with the time-independent Schroedinger equation

\begin{equation}
\frac{d^2}{dx^2} \phi(x) +\frac{2m}{\hbar^2}\biggl [ E-V(x) \biggr ] \phi(x) = 0,
\label{TISE}
\end{equation}
for a one-dimensional stepwise potential as defined in Eq. (\ref{Potential}).
Equation (\ref{TISE}) admits, for $E \geq 0$, a continuous and doubly degenerate spectrum. 
For each energy  eigenvalue $E$ there are two eigenstates $\phi_k(x)$ and 
$\phi_{-k}(x)$ with the positive wave vector $k$ defined as
\begin{equation}
k = \sqrt{\frac{2m}{\hbar^2} E}.
\end{equation}
By introducing the positive wave vectors in the regions at potentials $V_0$ and $V_1$
\begin{equation}
k_0 = \sqrt{ {2m \over \hbar^2} (V_0 - E)}, \hspace{0.5cm}
k_1 = \sqrt{ {2m \over \hbar^2} (E - V_1)},
\end{equation}
the eigenstates $\phi_k(x)$ can be expressed as
\begin{equation}
\phi_k(x)= (2 \pi)^{- {1\over 2}}
\begin{cases} e^{ikx} + r(k)e^{-ikx}        & x < a \\
        A^+(k) e^{k_0 x} + A^-(k) e^{-k_0 x}     & a \leq x < b \\
        B^+(k) e^{ik_1 x} + B^-(k) e^{-ik_1 x}   & b \leq x < c \\
        C^+(k) e^{k_0 x} + C^-(k) e^{-k_0 x}     & c \leq x < d \\
        t(k) e^{ikx}                             & x \geq  d.   \end{cases}
\end{equation}        
        
In this form the state $+k$ describes the stationary state of a particle coming from $x = -\infty$ with momentum $\hbar k$.
The normalization is chosen in such a way that the eigentates are orthogonalized with respect to the wave vector \cite{Landau1977}

\begin{equation}
\int_{-\infty}^{+\infty} dx~ \phi_k(x)^* \phi_{k'}(x)=\delta(k-k').
\end{equation}
The particle is reflected or transmitted respectively with probability 
\begin{equation}
R(k)=|r(k)|^2, \hspace{0.5cm} T(k)=|t(k)|^2.
\end{equation} 

The reflection and transmission amplitudes $r(k)$ and $t(k)$ are determined together with the 
coefficients $A^\pm(k)$, $B^\pm(k)$ and $C^\pm(k)$ by imposing the continuity of the eigenstates and 
their first derivatives in the discontinuity points of the potential. This leads to

\begin{equation}
t(k)={16 k k_1 k_0^2 \over f_-(k) - f_+(k)} e^{-ik(d-a)},
\end{equation}
where
\begin{eqnarray}
f_\pm(k)&=& e^{\mp k_0(b-a)} (k_0 \pm ik)\biggl [e^{-ik_1(c-b)} (k_0 \pm ik_1) g_+(k) + \nonumber \\
 & & e^{+ik_1(c-b)} (k_0 \mp ik_1) g_-(k) \biggr ],\\
g_\pm(k)&=&e^{\mp k_0(d-c)} (k_0 + ik_1)(k_0 \pm ik) - \nonumber \\
 & & e^{\pm k_0(d-c)} (k_0 - ik_1)(k_0 \mp ik).
\end{eqnarray}         
The remaining coefficients are then determined as
\begin{eqnarray}
A^\pm(k)= {e^{\mp k_0b} \over 2k_0}\biggl[B^+(k)e^{ ik_1b}(k_0 \pm ik_1) + \nonumber \\
B^-(k) e^{-ik_1b} (k_0 \mp ik_1)\biggr ],\\
B^\pm(k)= {e^{\mp ik_1c} \over 2ik_1}\biggl[C^+(k) e^{ k_0c} (ik_1 \pm k_0) + \nonumber \\ 
C^-(k) e^{-k_0c} (ik_1 \mp k_0) \biggr ], \\
C^\pm(k)= {e^{\mp k_0 d} \over 2k_0} t(k) e^{ikd} (k_0 \pm ik), \\
r(k)= {e^{ika} \over 2ik} \times \biggl [ A^+(k) e^{ k_0a} (ik - k_0) + \nonumber \\ A^-(k) e^{-k_0a} (ik + k_0) \biggr ].
\end{eqnarray}         

\section{Influence of the shape of the barrier on the convexity of the tunneling coefficient}%
\label{AppB}

In this appendix we show with a representative example how the shape of the barrier strongly influences the reactivity gain, even 
within the WKB approximation for which the only relevant quantity is the area of the classically forbidden region. 
In the case of a rectangular barrier of height $V_0$ and thickness $a$, with the particle mass and energy 
$m$ and $E$ respectively, the WKB approximation yields

\begin{equation}
T_r(E) \sim \exp{\left[-\frac{2a}{\hbar}\sqrt{2 m V_0}  \left(1 - \frac{E}{V_0} \right)^{1/2}\right]},
\end{equation}
which, in the limit $E/V_0 \rightarrow 0$, can be expanded as
\begin{equation}
T_r(E) \sim \left(1+ \frac{\sqrt{2 m V_0} a}{\hbar} \frac{E}{V_0} \right) \exp{\left(-\frac{2a}{\hbar} \sqrt{2 m V_0} \right)} ,
\label{Trectangular}
\end{equation}
a linear dependence on $E$ in the same limit, implying no initial curvature. 

If we instead consider a case with a smoother barrier, such as the following containing a Coulomb-like component

\begin{equation}
V(x)=\begin{cases} 
V_0 & x\in [-a,+a] \\ 
V_0~ a/|x| & x\in (-\infty,-a] \cup [+a,\infty),
\end{cases}
\end{equation}
we obtain, always in the WKB approximation
\begin{equation}
T_C(E) \sim \exp{\left[-\frac{2a}{\hbar} \sqrt{\frac{2 m V_0^2}{E}}  \arctan\left(\sqrt{\frac{V_0}{E}-1}\right)\right]},
\end{equation}
where $T_C(E)$ is observed to be convex in the entire range $0 \leq E/V_0 \leq 1$.  More specifically, in the $E/V_0 \rightarrow 0$ limit, the 
transmission coefficient is approximated as
\begin{equation} 
T_C(E) \sim \exp{\left(-\frac{\pi a}{\hbar} \sqrt{\frac{2 m V_0^2}{E}}\right)}.
\end{equation}
Notice the non-analytical dependence $T_C \propto \exp(-A/\sqrt{E})$, which implies a very soft increase 
and therefore a positive curvature at small values of $E$, i.e. $T_C(E)$ is convex. This is easily interpreted 
in terms of the behavior of the barriers as the energy of the impinging particle is increased. In the case of the rectangular 
barrier the increase in energy results in a linear decrease of the area of the classically forbidden region, therefore 
implying a square root dependence for the argument of the integral in the exponent of the WKB relationship. 
Instead, in the case of the Coulomb potential the decrease of the area of the classically forbidden region when 
increasing $E$ has a stronger dependence on $E$, at least initially. This creates convexity of $T(E)$ at low energy. 
Under these conditions, as discussed in Section IV, spreading the energy distribution can be advantageous for enhancing the reactivity.

\end{document}